\documentclass[twocolumn,showpacs,preprintnumbers,amsmath,amssymb]{revtex4-1}

\usepackage{color}
\usepackage{graphicx}
\usepackage{dcolumn}
\usepackage{bm}
\usepackage{comment}

\begin{document}

\preprint{}

\title{Electric dipole moment of $^{199}$Hg atom from P, CP-odd electron-nucleon interaction}

\newcommand{\saitama}{
Department of Physics, Saitama University, Saitama City 338-8570, Japan
}

\newcommand{\chiba}{
Department of Physics, Chiba Institute of Technology, Narashino, Chiba 275-0023, Japan
}

\newcommand{\ipno}{
IPNO, Universit\'{e} Paris-Sud, CNRS/IN2P3, F-91406, Orsay, France
}
\newcommand{\Riken}{
Theoretical Research Division, Nishina Center, RIKEN, Saitama 351-0198, Japan
}

\author{K. Yanase}
\email{yanase@nuclei.th.phy.saitama-u.ac.jp}
\affiliation{\saitama}

\author{N. Yoshinaga}
\email{yoshinaga@phy.saitama-u.ac.jp}
\affiliation{\saitama}

\author{K. Higashiyama}
\email{koji.higashiyama@it-chiba.ac.jp}
\affiliation{\chiba}

\author{N. Yamanaka}
\email{yamanaka@ipno.in2p3.fr}
\affiliation{\ipno}
\affiliation{\Riken}

\date{\today}

\begin{abstract}
We calculate the effect of the P, CP-odd electron-nucleon interaction on the electric dipole moment of the $^{199}$Hg atom by evaluating the nuclear spin matrix elements in terms of the nuclear shell model.
It is found that the neutron spin matrix element of the $^{199}$Hg nucleus is $\langle \Psi |\, \sigma_{nz} | \Psi \rangle \approx -0.4$ with a dominant configuration of $p_{1/2}$ orbital neutron.
We also derive constraints on the CP phases of Higgs-doublet models, supersymmetric models, and leptoquark models from the latest experimental limit $|d_{\rm Hg}| < 7.4 \times 10^{-30}e$ cm.
\end{abstract}


\maketitle

\section{\label{sec:intro}Introduction}

The baryon abundant Universe is realized if the fundamental theory fulfills Sakharov's criteria \cite{sakharov}.
An important issue is the insufficiency of CP violation in the standard model (SM) which generates a too small baryon number asymmetry \cite{farrar,huet}.
Motivated by this problem, the search for CP violation beyond the SM is nowadays an active area in particle physics.

One of the promising experimental approaches to unveil new sources of CP violation is the measurement of the {\it electric dipole moment} (EDM) \cite{hereview,bernreutherreview,jungmann,naviliatreview,khriplovichbook,ginges,pospelovreview,raidal,fukuyama,engeledmreview,yamanakabook,roberts,yamanakanuclearedmreview,atomicedmreview,chuppreview,safronova,orzel}.
The EDM is an excellent probe of CP violation thanks to the small SM background \cite{ckm,jarlskog}, and has been measured in many systems such as the paramagnetic atoms \cite{regan}, the neutron \cite{baker}, the muon \cite{muong2}, and the electron inside paramagnetic molecules \cite{hudson1,hudson2,acme,Andreev:2018ayy}.
There are also many new experimental ideas to measure the EDM in other systems such as the paramagnetic atoms using three-dimensional optical lattice \cite{chin,sakemi}, the proton and light nuclei using storage rings \cite{yamanakanuclearedmreview,storage1,storage2,storage4,Anastassopoulos,jedi2}, the strange and charmed baryons using bent crystals \cite{botella,Baryshevsky}, the $\tau$ lepton from the precision analysis of collider experimental data \cite{xinchen,koksal}, the electron in polar molecules and an inert gas matrix \cite{inertgasmatrix}, or in polyatomic molecules \cite{Kozyryev:2017cwq}, etc.
Currently, the most precise measurements are performed for the diamagnetic atoms.
They include $^{129}$Xe (current limit $|d_{\rm Xe}| < 4.81 \times 10^{-27} e\, {\rm cm}$ \cite{Sachdeva}, prospective sensitivity $\sim 10^{-31} e\, {\rm cm}$), $^{225}$Ra (current limit $|d_{\rm Ra}| < \times 10^{-23} e\, {\rm cm}$ \cite{bishof}, prospective sensitivity $\sim 10^{-28} e\, {\rm cm}$), and $^{199}$Hg which gives the current world record of the EDM upper limit $|d_{\rm Hg}| < 7.4 \times 10^{-30}e$ cm \cite{graner}.

Despite the exhaustive search for new physics in LHC experiments, their results are still consistent with the SM, and the development of other complementary approaches is now required.
In this regard, the EDM is actually an excellent alternative because it is more sensitive than the current reach of LHC (13 TeV) under the assumption of natural [$O(1)$] CP phases.
The sensitivity of the EDM is so high that the detection of the Lorentz violation \cite{Araujo:2015zsa,Araujo:2016hsn,araujo} or axionlike particle mediated interactions \cite{Mantry:2014zsa,stadnik} using this observable is also under discussion.

In the context of the atomic EDM, Schiff's screening phenomenon which states the cancellation of the EDM of nonrelativistic point-like constituents in electrically neutral bound states is well-known \cite{schiff}.
The EDM of atoms is generated by three important mechanisms. 
These are (a) the relativistic enhancement of the electron EDM, (b) the polarization of the atomic system by the P, CP-odd electron-nucleon ($e$-$N$) interaction, and (c) the nuclear Schiff moment giving the nonpointlike effect of the nucleus.
Among them, the atomic polarization due to the P, CP-odd $e$-$N$ interaction is a purely atomic effect, and it can only be studied by measuring the EDM of heavy atoms \cite{ginges,khriplovichbook,bouchiat,hinds,martensson,flambaumshellmodel2,dzuba2,barr2,hemckellar}.
The P, CP-odd $e$-$N$ interaction is recently attracting attention, since it is generated in the two-Higgs doublet model (2HDM) at the tree level, and has extensively been discussed in the literature \cite{barr3,hemackellarpakvasa,barr4,jung1}.
This process is also interesting due to its specific sensitivity to several classes of models such as the supersymmetric models with large $\tan \beta$ \cite{pospelovreview,fischler,Lebedev:2002ne,Demir:2003js,pilaftsisgluoniccpve-n,edmmssmreloaded,Lee:2012wa} or the $R$-parity violation \cite{herczege-n1,herczege-n2,cpve-n1loop,yamanakabook,rpvlinearprogramming}, as well as the leptoquark models \cite{barr2,hemackellarpakvasa,mahanta,herczegleptoquark,fuyuto} where the $s$-channel electron-quark interaction is relevant.
Since atoms are one of the most accurately measurable systems and give now the world record in the EDM experiments, the study of the contribution of the CP-odd $e$-$N$ interaction has an inevitable importance in the analysis of the CP violation of these models.

In the phenomenological analyses of the new physics beyond the SM using the atomic EDM, hadronic, nuclear and atomic level calculations are required.
In the context of the CP-odd $e$-$N$ interaction, the results of the calculations at the hadronic level are already reaching a very good accuracy thanks to the progress of lattice QCD, currently being precise at the level of 10\% \cite{chiqcdsigmaterm,rqcdsigmaterm,bmwsigmaterm,etmsigmaterm,green,rqcdisovector,Yamanaka:2015lfk,pndmeisovector,chiqcdisovector,etm2017,pndmetensor}.
The atomic level calculations are also known to be very accurate, and the theoretical uncertainty associated with them is typically of $O(1\% )$ \cite{martensson,dzuba,latha,singh,radziute,sahoo,sahoo2}.
For the nuclear level calculation of the EDM of $^{199}$Hg, however, the nuclear spin matrix elements, which are the most important input, have never been quantitatively evaluated.
An estimation using the simple nuclear shell model assuming only one valence nucleon is possible \cite{flambaumshellmodel1,flambaumshellmodel2,flambaumshellmodel3}, but it is known in modern nuclear physics that this na\"{i}ve picture does not hold in general due to the configuration mixing, in particular for nuclei with the nucleon numbers not close to the magic numbers.

For the case of the Xe atom, the nuclear level inputs required in the analysis of the atomic EDM such as the coefficients relating the intrinsic nucleon EDM and the CP-odd nuclear force to the nuclear Schiff moment or the nuclear spin matrix elements, were evaluated in the shell model with configuration mixing \cite{atomicedmreview,yoshinaga1,yoshinaga2,yoshinaga3,teruya}.
Combined with the results of atomic level calculations \cite{martensson,dzuba,lathaxerayb}, they allowed a quantitative access to the CP violation at the hadronic level and beyond.
Following the success of the nuclear level calculation of $^{129}$Xe, we expect 
the application of the same framework to $^{199}$Hg to have the best impact on the study of new physics contributing to the CP-odd $e$-$N$ interaction.
In this paper we therefore accurately evaluate the nuclear spin matrix elements of the $^{199}$Hg nucleus using the shell model, to allow quantitative analysis of theories contributing to the CP-odd $e$-$N$ interaction, i.e., the extended Higgs models, the supersymmetric models, and the leptoquark models.

This paper is organized as follows.
We first present in Section \ref{sec:CPVeN} the CP-odd $e$-$N$ interaction and its relation with the EDM of the $^{199}$Hg atom.
We then give the detail of the framework of the calculation of the spin matrix elements of $^{199}$Hg and the neighboring odd-nuclei in the shell model in Section \ref{sec:shellmodel}, and show the result as well as the relation between the quark-gluon level CP-odd interaction and the EDM of $^{199}$Hg. 
We then analyze in Section \ref{sec:constraint} the constraints on several candidates of new physics beyond the SM.
The last section is devoted to the summary.

\section{\label{sec:CPVeN}The P, CP-odd electron-nucleon interaction}

The leading P, CP-odd electron-nucleon ($e$-$N$) interaction contributing to the atomic EDM is given by three types of dimension-six contact interactions \cite{barr2,ginges,khriplovichbook,yamanakabook}:
\begin{eqnarray}
\hspace{-1em}
{\cal L}_{eN} 
&=&
 -\frac{G_F}{\sqrt{2}} \sum_{N=p,n} \Bigl[
C_N^{\rm SP} \bar NN \, \bar e i \gamma_5 e
+C_N^{\rm PS} \bar Ni\gamma_5 N \, \bar e e 
\nonumber\\
&& \hspace{6em}
-\frac{1}{2}C_N^{\rm T} \epsilon^{\mu \nu \rho \sigma} \bar N \sigma_{\mu \nu} N \, \bar e \sigma_{\rho \sigma} e 
\Bigr] .
\ \ \ \ \ 
\label{eq:pcpve-nint}
\end{eqnarray}
Here we denote the first, second, and third terms by the scalar-pseudoscalar (SP), the pseudoscalar-scalar (PS), and the tensor (T) type interactions, respectively.
Note the minus sign in front of $C_N^{\rm T}$ due to the convention.

The EDM of the $^{199}$Hg atom is given by the leading order perturbation of the CP-odd electron-nucleus interaction $H_{CP}$:
\begin{equation}
d_{\rm Hg}
=
2\sum_m
\frac{\langle \psi_0 | -e \sum_i {\mathbf r}_i | \psi_m \rangle \langle \psi_m | H_{CP} | \psi_0 \rangle  }{E_0 - E_m}
,
\end{equation}
where $\psi_m$ labels the atomic eigenstates. 
From the nonrelativistic spin structure of the CP-odd $e$-$N$ interaction (\ref{eq:pcpve-nint}), it is possible to parametrize the EDM of the $^{199}$Hg atom in the leading order of the CP-odd $e$-$N$ couplings as \cite{ginges}
\begin{eqnarray}
d_{\rm Hg}
&=&
{\cal R}_{\rm T} 
\Bigl( 
C_p^{\rm T}
\langle  \sigma_{pz} \rangle
+
C_n^{\rm T}
\langle \sigma_{nz} \rangle
\Bigr)
\nonumber\\
&&+
{\cal R}_{\rm PS} 
\Bigl( 
C_p^{\rm PS}
\langle  \sigma_{pz} \rangle
+
C_n^{\rm PS}
\langle \sigma_{nz} \rangle
\Bigr)
\nonumber\\
&&+
{\cal R}_{\rm SP} 
\Bigl( 
0.40 
C_p^{\rm SP}
+
0.60
C_n^{\rm SP}
\Bigr)
,
\end{eqnarray}
where $\langle \sigma_{Nz} \rangle \equiv \langle ^{199}{\rm Hg} | \,\sigma_{Nz} |^{199}{\rm Hg} \rangle$ $(N=p,n)$ is the nuclear spin matrix element.
We see that the T and PS type CP-odd $e$-$N$ interactions depend on the nuclear spin matrix elements, and we therefore need their accurate values to quantitatively analyze the effect of new physics.
The factors 0.40 and 0.60 are the fractions of the proton and neutron numbers over the total one, and are exact in the nonrelativistic limit.
For the case of the SP type CP-odd $e$-$N$ interaction, the calculation of nuclear matrix elements is therefore not required.

The atomic coefficient ${\cal R}_T$ was calculated in several atomic level approaches \cite{atomicedmreview,dzuba,latha,singh,radziute,sahoo,sahoo2}.
Here we use the result of the calculation in the relativistic normal coupled-cluster method ${\cal R}_T \approx -3.3 \times 10^{-20} e$ cm \cite{sahoo2}, where the theoretical uncertainty was estimated to be about 2\%.

The PS type CP-odd $e$-$N$ interaction is analytically related to the T type one with \cite{ginges,dzuba}
\begin{eqnarray}
\Bigl( C_p^{\rm PS} \langle \vec \sigma_p \rangle
+C_n^{\rm PS} \langle \vec \sigma_n \rangle 
 \Bigr)
&\leftrightarrow &
\frac{5 m_N R }{ Z \alpha_{\rm em} }  
\Bigl(
C_p^{\rm T} \langle \vec \sigma_p \rangle
+C_n^{\rm T} \langle \vec \sigma_n \rangle 
\Bigr)
,
\nonumber\\
\label{eq:cpsct}
\end{eqnarray}
where $R$ is the nuclear radius.
From this relation we have ${\cal R}_{\rm PS} \approx - 1.2 \times 10^{-22} e$ cm.

The $^{199}$Hg atom is a diamagnetic atom and has a closed electron shell, so the EDM is not generated by the SP type CP-odd $e$-$N$ interaction in the leading order of electromagnetic interaction.
It is, however, induced by the hyperfine interaction \cite{ginges,flambaumshellmodel2,fleig}.
The atomic level coefficient of the SP type CP-odd $e$-$N$ interaction ($C_N^{\rm SP}$) has recently been calculated, giving ${\cal R}_{\rm SP} \approx -2.8 \times 10^{-22}e$ cm \cite{fleig}, with about 20\% of uncertainty.

Let us move on to the derivation of the CP-odd $e$-$N$ interaction from the quark-gluon level physics.
The CP-odd $e$-$N$ couplings of Eq. (\ref{eq:pcpve-nint}) are generated by the CP-odd electron-quark interactions, which have the same Lorentz structure, and by the CP-odd electron-gluon interactions,
\begin{eqnarray}
\hspace{-1em}
{\cal L}_{eq/g}
&=&
-\frac{G_F}{\sqrt{2}} 
\Biggl\{
\sum_{q} \Bigl[ 
C_q^{\rm SP} \bar qq \, \bar e i \gamma_5 e+C_q^{\rm PS} \bar q i\gamma_5 q \, \bar e e 
\nonumber\\
&& \hspace{6em}
-\frac{1}{2}C_q^{\rm T} \epsilon^{\mu \nu \rho \sigma} \bar q \sigma_{\mu \nu} q \, \bar e \sigma_{\rho \sigma} e 
\Bigr] 
\nonumber\\
&&
+C_{eg}^{\rm SP} \frac{\alpha_s}{12\pi }G_{\mu \nu}^a G^{\mu \nu}_a \, \bar e i \gamma_5 e
+C_{eg}^{\rm PS} \frac{\alpha_s}{8\pi }\tilde G_{\mu \nu}^a G^{\mu \nu}_a \, \bar e e
\Biggr\}
,
\nonumber\\
\label{eq:electronquarkinteractions}
\end{eqnarray}
where $\tilde  G_{\mu \nu}^a \equiv \frac{1}{2} \epsilon_{\mu \nu \rho \sigma} G^{\rho \sigma}_a$.
Here note that $\alpha_s G_{\mu \nu}^a G^{\mu \nu}_a$ and $\alpha_s \tilde G_{\mu \nu}^a G^{\mu \nu}_a$ are renormalization group invariant in the leading order of the strong coupling $\alpha_s$.
The sum over the quark flavor $q$ has to be taken up to that allowed by the renormalization scale $\mu$ (e.g., $q=u,d,s,c$ for $\mu = 2$ GeV). 

For the T type CP-odd $e$-$N$ interaction, the couplings $C_N^{\rm T}$ are obtained by just multiplying the nucleon tensor charge $\delta q \equiv \frac{1}{2 m_N}
\langle p | \bar q i\sigma_{03} \gamma_5 q | p \rangle$ to the quark level analogue $C_q^{\rm T}$.
By writing them explicitly, we have 
\begin{eqnarray}
C_p^{\rm T}
&=&
C_u^{\rm T} \delta u +C_d^{\rm T} \delta d + C_s^{\rm T} \delta s+ C_c^{\rm T} \delta c
,
\\
C_n^{\rm T}
&=&
C_u^{\rm T} \delta d +C_d^{\rm T} \delta u + C_s^{\rm T} \delta s+ C_c^{\rm T} \delta c
,
\end{eqnarray}
where we assumed the isospin symmetry.
The light quark contributions to the nucleon tensor charge are given by $\delta u = 0.8$, $\delta d = -0.2$, and $\delta s = -0.003$ from recent lattice QCD calculations \cite{rqcdisovector,Yamanaka:2015lfk,pndmetensor,pndmeisovector,etm2017}, with 10\% of theoretical uncertainty for $\delta u, \delta d$, and 50\% for $\delta s$, being conservative.
The values of $\delta u$ and $\delta d$ are also extracted from experimental data using perturbative QCD based frameworks \cite{bacchetta,radici,kang,radici2}, but more accurate data with future experiments \cite{courtoy,yez,gaotensor,accardi} are required to perform quantitative analyses.
The charm quark contribution is consistent with zero within the uncertainty $|\delta c|< 0.005$ \cite{etm2017}.

We note that $C_q^{\rm T}$ $(q=u,d,s,c)$ calculated at the scale of the new physics (e.g., $\mu = 1$ TeV) has to be renormalized down to the scale where the results of the calculations of the nucleon tensor charges are available ($\mu = 2$ GeV in most of the cases).
The running of $C_q^{\rm T}$ from $\mu = 1$ TeV down to $\mu = 2$ GeV brings a factor of about 0.8 \cite{barone,artru,degrassi,renormalizationedm,renormalizationedm2}.
This suppression is intuitively understood by the radiative emission and absorption of the gluon, which flip the spin of the quark.
A similar mechanism is also working at the level of the nucleon matrix element, attenuating the contribution of the quark to the nucleon tensor charge \cite{tensorsde,pitschmann}.

For the PS type CP-odd $e$-$N$ interaction, we have to consider the gluonic contribution in addition to the CP-odd electron-quark interaction.
The couplings are given by 
\begin{eqnarray}
C_p^{\rm PS}
&=&
C_u^{\rm PS} \langle p | \bar u i \gamma_5 u | p \rangle 
+C_d^{\rm PS} \langle p | \bar d i \gamma_5 d | p \rangle 
\nonumber\\
&&
+ \hspace{-0.5em} \sum_{q'=s,c} \hspace{-0.5em} 
C_{q'}^{\rm PS} \langle p | \bar q' i \gamma_5 q' | p \rangle
+ C_{eg}^{\rm PS} \frac{\alpha_s}{8 \pi} \langle p | \tilde G_{\mu \nu}^a G^{\mu \nu}_a | p \rangle
,
\nonumber\\
\\
C_n^{\rm PS}
&=&
C_u^{\rm PS} \langle n | \bar d i \gamma_5 d | n \rangle 
+C_d^{\rm PS} \langle n | \bar u i \gamma_5 u | n \rangle 
\nonumber\\
&&
+ \hspace{-0.5em} \sum_{q'=s,c} \hspace{-0.5em}
C_{q'}^{\rm PS} \langle n | \bar q' i \gamma_5 q' | n \rangle
+ C_{eg}^{\rm PS} \frac{\alpha_s}{8 \pi} \langle n | \tilde G_{\mu \nu}^a G^{\mu \nu}_a | n \rangle
,
\nonumber\\
\end{eqnarray}
where the isospin symmetry was used.
The nucleon matrix elements are phenomenologically derived from the anomalous Ward identity \cite{herczege-n1,herczege-n2,cpve-n1loop,yamanakabook,alonso,cheng1,chengli,dienes,cheng2,scopel}.
Using the latest QCD level inputs \cite{pdg,fermilablattice,Campos:2018ahf}, we have $\langle p | \bar u i \gamma_5 u | p \rangle = 180$, $\langle p | \bar d i \gamma_5 d | p \rangle = -170$, $\langle N | \bar s i \gamma_5 s | N \rangle = -5.1$, $\langle N | \bar c i \gamma_5 c | N \rangle = -0.34$, and $\frac{\alpha_s}{8\pi} \langle N | \tilde G_{\mu \nu}^a G^{\mu \nu}_a | N \rangle = 400$ MeV ($N=p,n$), with about 30\% of theoretical uncertainty \cite{atomicedmreview}.
Here we note that the quark pseudoscalar matrix elements $\langle N | \bar q i \gamma_5 q | N \rangle$ are dimensionless (the same applies for quark scalar matrix elements $\langle N | \bar q q | N \rangle$).
It is important to note that the PS type CP-odd $e$-$N$ interaction is enhanced, due to the pion pole contribution of the light quarks \cite{axialsde,alonso}.
We also point out that at a scale below the quark mass threshold, the $b$ and $t$ quarks contribute to $C_{eg}^{\rm PS}$ through quark loops, as $C_{eg}^{\rm PS} = -\frac{1}{m_Q}C_{Q}^{\rm PS}$ ($Q = b,t$), in the leading order in $\alpha_s$ \cite{witten,shifman,hatsuda1,hatsuda2,Zhitnitsky}.
For the case of $C_{q}^{\rm PS}$, the renormalization grows the couplings when the scale decreases since the product of $\langle N | \bar q i \gamma_5 q | N \rangle$ and the current quark mass forms invariants.
The running from $\mu=1$ TeV to $\mu=2$ GeV yields an enhancement of a factor of about two (this is also valid for $b$ and $t$ quark contributions since $C_{eg}^{\rm PS} \propto \frac1{m_Q}$).

The SP type CP-odd $e$-$N$ interaction is derived in a similar way as the PS one:
\begin{eqnarray}
C_p^{\rm SP}
&=&
\frac{C_u^{\rm SP}}{2} \Bigl( \frac{\sigma_{\pi N}}{m_u+m_d} +g_S \Bigr) 
+\frac{C_d^{\rm SP}}{2} \Bigl( \frac{\sigma_{\pi N}}{m_u+m_d} -g_S \Bigr) 
\nonumber\\
&&
+ \hspace{-0.5em} \sum_{q'=s,c} \hspace{-0.5em}
C_{q'}^{\rm SP} \frac{\sigma_{q'}}{m_{q'}} 
+ C_{eg}^{\rm SP}  \frac{\alpha_s}{12 \pi} \langle p | G_{\mu \nu}^a G^{\mu \nu}_a | p \rangle
,
\\
C_n^{\rm SP}
&=&
\frac{C_u^{\rm SP}}{2} \Bigl( \frac{\sigma_{\pi N}}{m_u+m_d} -g_S \Bigr) 
+\frac{C_d^{\rm SP}}{2} \Bigl( \frac{\sigma_{\pi N}}{m_u+m_d} +g_S \Bigr) 
\nonumber\\
&&
+ \hspace{-0.5em}\sum_{q'=s,c} \hspace{-0.5em} 
C_{q'}^{\rm SP} \frac{\sigma_{q'}}{m_{q'}}
+ C_{eg}^{\rm SP}  \frac{\alpha_s}{12 \pi} \langle n | G_{\mu \nu}^a G^{\mu \nu}_a | n \rangle
, \ \ \ \ \ 
\end{eqnarray}
where we again used the isospin symmetry.
Here the pion-nucleon sigma term is given by $\sigma_{\pi N} \approx 40$ MeV \cite{chiqcdsigmaterm,rqcdsigmaterm,bmwsigmaterm,etmsigmaterm} and the isovector nucleon scalar charge is $g_S \approx 0.9 $ at the scale $\mu = 2$ GeV \cite{rqcdisovector,pndmeisovector,chiqcdisovector,green,etm2017}, with errorbars of about 10\%.
We note here that $\sigma_{\pi N}$ phenomenologically extracted from experimental data is around 60 MeV \cite{alarcon1,alarcon2,Hoferichter,Hoferichter2,yao,deelvira}.
The strange and charm contents of the nucleon are given by $\sigma_{s} \approx 40$ MeV \cite{chiqcdsigmaterm,rqcdsigmaterm,bmwsigmaterm,etmsigmaterm,ohki1,ohki2,ohki3,alarcon3} and $\sigma_{c} \approx 80$ MeV \cite{lattice_charm_content1,lattice_charm_content2,hobbs}, respectively, with an uncertainty of about 50\%.
The gluonic condensate is derived from the trace anomaly as $\frac{\alpha_s}{12\pi} \langle N | G_{\mu \nu}^a G^{\mu \nu}_a | N \rangle = (-50 \pm 5)$ MeV ($N=p,n$) \cite{atomicedmreview}.
As with the PS type one, the contribution from $b$ and $t$ quarks at a scale below the threshold is given by $C_{eg}^{\rm SP} = -\frac1{m_Q}C_{b}^{\rm SP}$.
The renormalization of $C_{q/eg}^{\rm SP}$ is exactly the same as that of $C_{q/eg}^{\rm PS}$, since the product of $\langle N | \bar q q | N \rangle$ or $1/m_Q$ ($Q=b,t$) with the current quark mass is invariant under the renormalization group evolution.

\section{\label{sec:shellmodel}Nuclear shell-model calculation}

The nuclear shell model is one of the most successful models to describe various properties of nuclear structure such as the energy spectra, electromagnetic transitions, and electromagnetic moments.
In order to study the nuclear structure and estimate the nuclear spin matrix elements of $^{199}$Hg, it is preferable to exploit all the single-particle levels between magic numbers $82$ and $126$ for neutrons and those between magic numbers $50$ and $82$ for protons.
It is, however, very hard to perform the full shell-model calculations for $^{199}$Hg due to the large number of shell-model configurations.
To get around this problem, the pair-truncated shell model (PTSM) is utilized, where the full shell-model space is truncated within subspaces composed of collective pairs of nucleons.
The details of the PTSM framework are given in Refs.~\cite{Higashi03,Yoshi04,Higashi11-1,Higashi11-2}.

\begin{table}[h]
\centering
\caption{Single-particle energies 
$\varepsilon_{n}$ and $\varepsilon_{p}$ for neutron holes and proton holes, respectively, in units of MeV.
}
\label{tab-SPE}
\begin{ruledtabular}
\begin{tabular}{ccccccc}
 & $2p_{1/2}$	& $1f_{5/2}$	& $2p_{3/2}$	& $0i_{13/2}$	& $1f_{7/2}$	& $0h_{9/2}$ \\\hline
 $\varepsilon_{n}$
 & 0.000		& 0.570			& 0.898			& 1.633			& 2.340			& 3.415
 \\
 & $2s_{1/2}$	&$1d_{3/2}$		& $0h_{11/2}$	& $1d_{5/2}$	& $0g_{7/2}$	& \\\hline
 $\varepsilon_{p}$
 & 0.000		& 0.351			& 1.348			&  1.683		& 2.500		&
\end{tabular}
\end{ruledtabular}
\end{table}

\begin{table}[h]
\centering
\caption{Optimized strengths of two-body interactions between neutrons ($n$-$n$), 
protons ($p$-$p$), and neutrons and protons ($n$-$p$). $G_0$ and $G_2$
indicate the strengths of the monopole-pairing and quadrupole-pairing 
(QP) interactions between like nucleons.
$\kappa_2$ indicates the strengths of the quadrupole-quadrupole (QQ) interactions between like and unlike nucleons.
The strengths of QP and QQ interactions are given in units of $\text{MeV} / b^4$.
The oscillator parameter $b = \sqrt{ \hbar / M \omega }$ is determined by the nucleon mass $M$ and the frequency $ \hbar \omega = 41\, A^{-1/3}$ MeV.
}
\label{tab-PARA}
\begin{ruledtabular}
\begin{tabular}{cccc}
                  & $G_0$ & $G_2$  &$\kappa_2$ \\\hline
~${n}$-${n}$~ &0.080 & 0.006 & 0.016 \\
~${p}$-${p}$~ &0.080 & 0.006 & 0.004 \\
~${n}$-${p}$~ &       &        & 0.040
\end{tabular}
\end{ruledtabular}
\end{table}

For the single-particle orbitals in the shell-model space, we employ the six orbitals between magic numbers $82$ and $126$ ($0h_{9/2}$, $1f_{7/2}$, $0i_{13/2}$, $2p_{3/2}$, $1f_{5/2}$, and $2p_{1/2}$) for neutrons and the five orbitals between magic numbers $50$ and $82$ ($0g_{7/2}$, $1d_{3/2}$, $1d_{5/2}$, $2s_{1/2}$, and $0h_{11/2}$) for protons.
The single-particle energies listed in Table~\ref{tab-SPE} are determined to reproduce the low-lying spectra in $^{207}$Pb and $^{207}$Tl.
Both neutrons and protons are treated as holes.
The full shell-model space is truncated into subspaces composed of collective pairs with angular momentum zero ($S$ pairs), two ($D$ pairs), and four ($G$ pairs) for protons, and $S$ and $D$ pairs for neutrons.

As a nuclear effective interaction, the so-called monopole plus quadrupole-quadrupole interaction is employed~\cite{Higashi11-1,Higashi11-2}.
The Hamiltonian is written as
\begin{align}
\hat{H} = \hat{H}_{n} + \hat{H}_{p} + \hat{H}_{np} ,
\label{eq-eff-Ham}
\end{align}
where $\hat{H}_{n}$ and $\hat{H}_{p}$ consist of the single-particle energies and the two-body interactions for neutrons and protons, respectively, and $H_{np}$ represents the quadrupole-quadrupole interaction between a neutron and a proton.
The optimized strength parameters of the effective two-body interactions in Eq.~(\ref{eq-eff-Ham}), which are listed in Table~\ref{tab-PARA}, are determined by performing a $\chi^{2}$-fit to the experimental spectra for some low-lying states in $^{200}$Hg, $^{198}$Hg, $^{198}$Pt, and $^{196}$Pt nuclei.

We calculate odd-mass Hg isotopes with the use of the Hamiltonian in Eq.~(\ref{eq-eff-Ham})
and the strengths shown in Table~\ref{tab-PARA}.
The results of the nuclear spin matrix elements in the lowest $1/2^-$ state of $^{199}$Hg are 
\begin{align}
\bigl\langle \sigma_{nz} \bigr\rangle
 & \equiv \left\langle {\Psi \left| {{\sigma _{nz}}} \right|\Psi } \right\rangle  = - 0.3765 , \\
\bigl\langle \sigma_{pz} \bigr\rangle
 & = 0.0088,
\end{align}
where $ \left| \Psi  \right\rangle $ represents the wavefunction of the lowest $1/2^-$ state.

\begin{table}[h]
\renewcommand\arraystretch{1.4}
\centering
\caption{Nuclear spin matrix elements for odd-mass Hg isotopes.
}
\label{tab-spin-results}
\begin{ruledtabular}
\begin{tabular}{ccccc}
	& $^{199}$Hg	& $^{201}$Hg	& $^{203}$Hg	& $^{205}$Hg	\\ \hline
 $\bigl\langle \sigma_{nz} \bigr\rangle$
	& $-0.3765$	& $-0.3396$	& $-0.3494$	& $-0.3333$	\\
 $\bigl\langle \sigma_{pz} \bigr\rangle$
	& $0.0088$		& $-0.0152$	& $-0.0255$	& $-0.0016$
\end{tabular}
\end{ruledtabular}
\end{table}

The nuclear spin matrix elements of $^{199}$Hg are given as
$\bigl\langle \sigma_{nz} \bigr\rangle = -\frac{1}{3}$ and $\bigl\langle \sigma_{pz} \bigr\rangle = 0$
in the simple picture that an odd neutron resides in the $2p_{1/2}$ orbital~
\cite{ginges,flambaumshellmodel2,yamanakabook}.
Our results for $\bigl\langle \sigma_{nz} \bigr\rangle$ and $\bigl\langle \sigma_{pz} \bigr\rangle$ are consistent with the simple estimate.
Table~\ref{tab-spin-results} shows nuclear spin matrix elements of odd-mass Hg isotopes.
The results gradually approach the simple estimate as the number of valence neutrons decreases
from $^{199}$Hg to $^{205}$Hg.
Thus, it is concluded that the effect of the configuration mixing on the nuclear spin matrix elements is limited.

\begin{figure}[h]
\centering
\includegraphics[width=\linewidth]{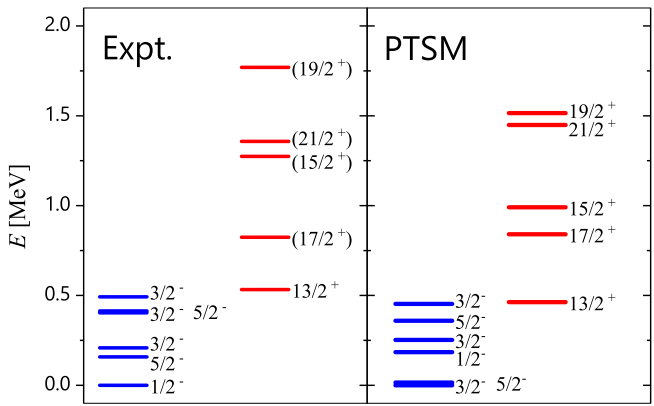}
\caption{\label{fig-Hg199spectrum}
The energy spectrum for $^{199}$Hg is given on the right panel (PTSM).
The optimized strength parameters of the effective two-body interactions are shown in Table\,\ref{tab-PARA}.
The experimental data (Expt.) on the left panel is taken from Ref.~\cite{ENSDF}.
}
\end{figure}

Figure~\ref{fig-Hg199spectrum} shows the low-lying energy levels of $^{199}$Hg
in comparison to experimental data.
The actual spin-parity of the ground state is $1/2^-$,
whereas the lowest $1/2^-$ state is calculated at $0.185$~MeV in the present framework.

\begin{table}
\renewcommand{\arraystretch}{1.5}
\centering
\caption{
Magnetic dipole moments ${\mu}$ (Calc.) in units of the nuclear magneton ${\mu_N}$ in comparison with the experimental data (Expt.)~\cite{ENSDF,NDS205,NDS203,NDS201,NDS199}.
The lowest state for each spin-parity $I^{ \pi }$ and the second lowest one are indicated with $I^{ \pi }_1$ and $I^{ \pi }_2$, respectively.
}
\label{tab-mag-Hg}
\begin{ruledtabular}
\begin{tabular}{cccc}
			 	& $I^{ \pi }_i$ 		& Expt.			& Calc.			\\\hline
$^{205}$Hg 	& $ 1/2^-_1 $ 	& +0.601 		& +0.372 		\\
$^{203}$Hg 	& $ 5/2^-_1 $		& +0.849 		& +0.899 		\\
$^{201}$Hg 	& $ 3/2^-_1 $ 	& $-$0.560 	& $-$0.605 	\\
$^{199}$Hg 	& $ 1/2^-_1 $ 	& +0.506 		& +0.495 		\\
				& $ 5/2^-_1 $ 	& +0.88(3) 		& +0.748 		\\
				& $ 3/2^-_1 $ 	& $-$0.56(9) 	& $-$0.530 	\\
				& $ 5/2^-_2 $ 	& +0.80(9) 		& +1.000 		\\
				& $ 13/2^+_1 $ 	& $-$1.015 	& $-$1.080 	\\
\end{tabular}
\end{ruledtabular}
\end{table}

In order to examine the wavefunctions further,
nuclear magnetic moments of Hg isotopes are calculated.
The third component of the magnetic moment operator is defined in units of the nuclear magneton $\mu_{N}$ as
\begin{align}
 \hat{ \mu }_{ z } = \mu_{ N } \sum_{ \tau = n, p }
 \left(
  g_{ l \tau }^{ ( \text{eff} ) } \hat{ l }_{ \tau z }
  + g_{ s \tau }^{ ( \text{eff} ) } \hat{ s }_{ \tau z }
 \right) ,
\end{align}
where $\hat{ l }_{ \tau z }$ and $\hat{ s }_{ \tau z }$ stand for the third components of the orbital angular momentum operator and the spin operator, respectively.
In this study effective $g$-factors of $g_{ s \tau }^{ ( \text{eff} ) } = 0.6~g_{ s \tau }^{ ( \text{free} ) }$ are adopted,
where $g_{ s \tau }^{ ( \text{free} ) }$ indicates the free spin $g$-factors.
The quenching factors are consistent with the adjusted values in the large-scale shell-model study around the mass number 210 \cite{teruya2,yanase}.
As shown in Table~\ref{tab-mag-Hg},
the magnetic moments of odd-mass Hg isotopes can be systematically reproduced in our framework
except for $^{205}$Hg, which has only one valence neutron.

The quenching of the $g$-factors is due to the core-polarization effects
and the meson-exchange currents.
As discussed in Sec.\,3.4 of Ref.\,\cite{towner}, the core-polarization effects are differently affective between nuclei with just one valence nucleon and those with more than two valence nucleons because of the Pauli blocking.
On the other hand, the main part of the so-called core polarization
is the configuration mixing within the one harmonic-oscillator shell.
The configuration mixing is taken into account in the present PTSM framework
except for the neutron part of $^{205}$Hg.
In contrast, the contributions of the meson-exchange currents are expected to be independent of the number of valence nucleons~\cite{barzakh,chemtob}.
According to the above arguments on the origin of the quenching of the $g$-factors,
it is reasonable to employ different effective $g$-factors between $^{205}$Hg and $^{203,\,201,\,199}$Hg.
In other words, the PTSM framework has the ability to accurately predict the spin dependent quantities of $^{199}$Hg.

\begin{figure}[htb]
\begin{center}
\includegraphics[width=8cm]{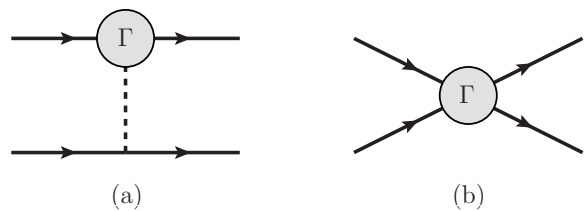}
\caption{\label{fig:2-body_correction}
The next-to-leading order two-nucleon correction to the nucleon operators, with (a) the pion-exchange current and (b) the two-nucleon contact interaction.
The solid and dashed lines indicate the nucleon and pion, respectively.
The scalar, pseudoscalar, and tensor operators are denoted by $\Gamma$.
}
\end{center}
\end{figure}

To complete the investigation of errorbars, we also comment on the next-to-leading order (NLO) contribution in the chiral effective field theory.
At the NLO, the nuclear scalar, tensor, and pseudoscalar matrix elements are corrected by the one-loop level pion dressing and the two-nucleon contribution, composed of the pion-exchange current and the short-range contact term (see Fig. \ref{fig:2-body_correction}).
In any cases, it can be shown from the power counting argument \cite{weinberg,park} that the above corrections are suppressed by at least a factor of $O(m_\pi^3/m_N^3) \simeq 0.3 \%$, thus being negligible in our analysis.
We note that the nuclear matrix elements $\langle \Psi |\, \sigma_{N} | \Psi \rangle$ are not suppressed by a factor of nucleon momentum and, consequently, the NLO contribution is relatively small.
The above discussion also applies to the correction of the CP-odd interactions due to the change of the model space, since the only source of the modification of the nucleon is the interaction either with other nucleons or with the virtual pions.
We can thus use the same CP-odd operators as the bare one [Eq. (\ref{eq:pcpve-nint})] to a good approximation.
In contrast, the magnetic moment is sizably corrected by the pion-exchange current, due to the momentum suppression which is present even at the tree level \cite{towner}.

The above results provide us the relation between $d_{\rm Hg}$ and the CP-odd $e$-$N$ couplings
\begin{eqnarray}
d_{\rm Hg}
&=&
\Bigl(
1.2 C_n^{\rm T}
+0.0045 C_n^{\rm PS}
\nonumber\\
&& \hspace{0.5em}
-0.017 C_n^{\rm SP}
-0.011 C_p^{\rm SP}
\Bigr)
\times 10^{-20}e\, {\rm cm}
.
\end{eqnarray}
We neglected the effect from the proton spin matrix element since it is small and its theoretical uncertainty exceeds the central value.
We also display the relation between $d_{\rm Hg}$ and the CP-odd quark or gluon level interactions defined in Eq. (\ref{eq:electronquarkinteractions}):
\begin{eqnarray}
d_{\rm Hg}
&=&
\Bigl(
-0.23 C_u^{\rm T}
+0.83 C_d^{\rm T}
-0.003 C_s^{\rm T}
\nonumber\\
&& \hspace{0.5em}
-1.4 C_u^{\rm PS}
+1.5 C_d^{\rm PS}
-0.04 C_s^{\rm PS}
-0.003 C_c^{\rm PS}
\nonumber\\
&& \hspace{0.5em}
-8\times 10^{-4} C_b^{\rm PS}
-2\times 10^{-5} C_t^{\rm PS}
+3.2 \, {\rm MeV}^{-1} C_{eg}^{\rm PS}
\nonumber\\
&& \hspace{0.5em}
-0.14 C_u^{\rm SP}
-0.15 C_d^{\rm SP}
-0.02 C_s^{\rm SP}
-0.003 C_c^{\rm SP}
\nonumber\\
&& \hspace{0.5em}
-6 \times 10^{-4} C_b^{\rm SP}
-2 \times 10^{-5} C_t^{\rm SP}
\nonumber\\
&& \hspace{4em}
+2.4 \, {\rm MeV}^{-1} C_{eg}^{\rm SP}
\Bigr)
\times 10^{-20}e\, {\rm cm}
,
\end{eqnarray}
with a conservative errorbar of 40\% for the $u$- and $d$-quark contributions, and 60\% for the others.
Note that the CP-odd electron-quark or gluon couplings were renormalized at $\mu = 1$ TeV.

\section{\label{sec:constraint}Constraints on new physics beyond standard model}


\begin{figure}[htb]
\begin{center}
\includegraphics[width=9cm]{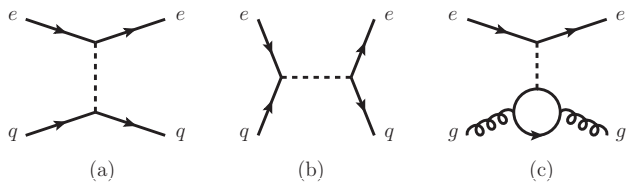}
\caption{\label{fig:diagrams}
Diagramatic representation of the leading contribution of new physics to the CP-odd $e$-$N$ interaction.
(a) $t$-channel electron-quark interaction, (b) $s$-channel electron-quark interaction, (c) electron-gluon interaction.
For (b), the dashed line of the intermediate state may be a scalar or a vector boson.
For (c), the particle in the loop can be a quark as well as a colored scalar field.
}
\end{center}
\end{figure}

Let us now discuss concrete constraints for several known models.
The first candidate to be considered is the 2HDM.
In the standard 2HDM, there are five physical Higgs modes, of which three are charge neutral including a CP-odd one.
The relevant lagrangian is the Yukawa interaction between the Higgs bosons and the SM fermions:
\begin{eqnarray}
{\cal L}_Y
&=&
-\sum_{i=e,d,s,b}
\frac{Y_i}{\sqrt{2}}
\Bigl[
(
\sin \beta \, \phi^0_u 
+\cos \beta \, \phi^0_d 
)
\bar \psi_i \psi_i
\nonumber\\
&&\hspace{10em}
+
\cos \beta \, 
A \bar \psi_i i \gamma_5 \psi_i
\Bigr]
\nonumber\\
&&
-\sum_{i=u,c,t}
\frac{Y_i}{\sqrt{2}}
\Bigl[
(
\cos \beta \, \phi^0_u 
-\sin \beta \, \phi^0_d 
)
\bar \psi_i \psi_i
\nonumber\\
&&\hspace{10em}
-
\sin \beta \, 
A \bar \psi_i i \gamma_5 \psi_i
\Bigr]
,
\end{eqnarray}
where $\phi^0_u, \phi^0_d$, and $A$ are the CP-even up-type, CP-even down-type, and the CP-odd Higgs fields, respectively, and $Y_i$ denotes the Yukawa coupling of the fermion $i$.
The parameter $\beta$ is defined by the well-known ratio between the vacuum expectation values of the up- and down-type Higgs fields $\tan \beta \equiv \frac{v_u}{v_d}$.
Here we note that the CP violation appears in the mixing between two Higgs modes with opposite CP parity \cite{tdlee,weinbergoperator,gunionreview,branco,ivanov,YongchaoZhang,ruisantos}.
The physical state with the lowest mass is identified as the Higgs boson with mass $m_H=125$ GeV \cite{higgs1,higgs2,higgs3}.

This mixing generates SP and PS type CP-odd $e$-$N$ interactions at the tree level [see Fig. \ref{fig:diagrams} (a)], as \cite{Weinberg:1990me,Kao:1992jv}
\begin{eqnarray}
-\frac{G_F}{\sqrt{2}}C_j^{\rm SP}
&=&
-\frac{Y_e Y_j }{ 4m_H^2} \cos \beta \sin \beta \tan^2 \beta \, {\rm Im }(\tilde Z_2)
,
\\
-\frac{G_F}{\sqrt{2}}C_k^{\rm SP}
&=&
\frac{Y_e Y_k }{ 4m_H^2}\cos^2 \beta \, {\rm Im }(\tilde Z_1)
,
\\
-\frac{G_F}{\sqrt{2}}C_j^{\rm PS}
&=&
-\frac{Y_e Y_j }{ 4m_H^2}
\sin \beta \cos \beta \cot^2 \beta \, {\rm Im }(\tilde Z_1)
,
\\
-\frac{G_F}{\sqrt{2}}C_k^{\rm PS}
&=&
\frac{Y_e Y_k }{ 4m_H^2} \cos^2 \beta \, {\rm Im }(\tilde Z_1)
,
\end{eqnarray}
where $j=u,c,t$ and $k=d,s,b$.
We note that $m_j = Y_j \sin \beta / \sqrt{2\sqrt{2}G_F} $ ($j=u,c,t$) and $m_k = Y_k \cos \beta / \sqrt{2\sqrt{2}G_F} $ ($k=e,d,s,b$).
The dimensionless constants $\tilde Z_1$ and $\tilde Z_2$ depend on the detail of the Higgs sector, especially on the mass of the CP-odd Higgs mode $A$.
The mass of $A$ is undetermined, but constrained to be $m_A > 1.5$ TeV in the large $\tan \beta$ scenario \cite{Aaboud:2017sjh,Sirunyan:2018zut}.
Their dependence on $\beta$ is factorized as $\tilde Z_1 \propto \tan \beta \sqrt{1+\tan^2 \beta} $ and $\tilde Z_2 \propto \cot \beta \sqrt{1+\cot^2 \beta} $ (for the convention, see Ref. \cite{Weinberg:1990me}).
In the case where $\tan \beta$ is large, 
the experimental data of $d_{\rm Hg}$ then yield the following bound:
\begin{eqnarray}
\bigl| {\rm Im} (\tilde Z_1)  \bigr| 
&<&
1.8
,
\label{eq:2hdmlimitb}
\end{eqnarray}
where we also considered the systematic errors.
We note that ${\rm Im} (\tilde Z_1)$ grows as $\tan^2 \beta$ in the 2HDM, so that the constraint on the Higgs sector becomes significantly tight for large $\tan \beta$.
In the above discussion, we did not constrain the CP violation of the top quark sector \cite{ellishwang,demartin,Chien:2015xha,cirigliano,cirigliano2,mileo,englert,coleppa,Fuyuto:2017xup,cpvhiggseleposi,Brandstetter:2018eju,goncalves} because it is more tightly bounded by the analysis of the Barr-Zee type diagram \cite{barr-zee,gunion,gunion2,leigh,completebarr-zee,muonbarr-zee,buras2hdmedm,brod,tabe,inoue2hdmedm,jung1,higgcisionedm,dekens,shuminzhao,Kobakhidze,balazs,nakai,haolinli,Panico,Bruggisser1,Huang,Bruggisser2}.

The supersymmetric models \cite{mssm1,mssm2,mssm3} include the 2HDM as the limit of heavy sparticles.
If the sfermions are relevant, additional contributions to the CP phase of the Yukawa coupling from loop diagrams are possible \cite{bingli,carena,Heinemeyer1,passehr,borowka,Heinemeyer2}.
The first effect we might consider is the squark loop inducing the Higgs-gluon-gluon ($H G_{\mu \nu ,a} G_a^{\mu \nu}$) vertex which contributes to the SP type CP-odd $e$-$N$ interaction [see Fig. \ref{fig:diagrams} (c)] \cite{pilaftsisgluoniccpve-n}.
This contribution is, however, well below that given by the CP-odd electron-quark interaction, as we will see below.

The correction to the Yukawa vertex with intermediate sfermion propagation may be important in the case of large $\tan \beta$.
This effect actually brings an additional $\tan \beta$ to the CP-odd $e$-$N$ interaction, resulting in a $\tan^3 \beta$ dependence \cite{Lebedev:2002ne,Demir:2003js}.
For large $\tan \beta$, we actually have
\begin{eqnarray}
\Biggl| \frac{ 
\tan^2 \beta (\sin \delta_e -\sin \delta_d ) }{
(1-J_e \tan \beta) (1-J_d \tan \beta) } \Biggr| 
&<&
1.8
,
\\
\Biggl| \frac{
\tan^2 \beta (\sin \delta_e -\sin \delta_s ) }{
(1-J_e \tan \beta) (1-J_s \tan \beta) } \Biggr| 
&<&
1.9
,
\\
\Biggl| \frac{
\tan^2 \beta (\sin \delta_e -\sin \delta_b ) }{
(1-J_e \tan \beta) (1-J_b \tan \beta) } \Biggr| 
&<&
2.0
,
\ \ \ \ \ \ \ \ 
\end{eqnarray}
where $\sin \delta_i = \frac{{\rm Im}(J_i) \tan \beta }{|1+J_i \tan \beta|} $ ($i=e,d,s,b$).
The above equations show almost no dependence on flavor, despite the hierarchy between the Yukawa couplings.
They are actually cancelled by the nucleon pseudoscalar density $\langle N|\bar q i \gamma_5 q|N\rangle$ which is approximately proportional to the inverse of the quark mass \cite{atomicedmreview}.
We note that the above bounds do not disappear in the limit of heavy supersymmetric particles \cite{Lebedev:2002ne,Demir:2003js,pospelovreview}.
For the case of the third generation, the two-loop level contribution to the electron EDM may be large \cite{pospelovreview,Lebedev:2002ne,Chang:1998uc,Pilaftsis:1999td,Chang:1999zw,Pilaftsis:2002fe,Demir:2003js,Yamanaka:2012ia}, although it decouples for heavy sparticle masses. 
If the sparticles have masses of $O(1)$ TeV (the current experimental lower limits of the masses of top squarks and gluinos are $m_{\tilde t} > 1$ TeV and $m_{\tilde g} > 2$ TeV, respectively \cite{Sirunyan:2017leh,Aaboud:2018kya,Aaboud:2018mna}), a stronger constraint by two or three orders of magnitude is obtained by the recent experimental data $d_e < 1.1 \times 10^{-29}e$ cm \cite{Andreev:2018ayy}.
Note that for the 2HDMs, interesting cancellations or fine-tunings may occur \cite{cancellationhiggs1,cancellationhiggs2}, although we do not consider them in this work.


In the supersymmetric extension of the SM, $R$-parity violating (RPV) interactions \cite{rpv1,rpv2,rpv3,rpv4}, which do not conserve lepton or baryon numbers, are also allowed.
The relevant RPV interactions are given by
\begin{eqnarray}
{\cal L }_{\rm R\hspace{-.4em}/\,} 
&=&
- \sum_{i,j,k}
\Biggl\{
\frac{ \lambda_{ijk}}{2} \Bigl[
\tilde \nu_i \bar e_k P_L e_j 
+\tilde e_{Lj} \bar e_k P_L \nu_i 
\nonumber\\
&& \hspace{6em}
+ \tilde e_{Rk}^\dagger \bar \nu_i^c P_L e_j 
-(i \leftrightarrow j ) \Bigr] 
\nonumber\\
&&+\lambda'_{ijk} \Bigl[
\tilde \nu_i \bar d_k P_L d_j 
+ \tilde d_{Lj} \bar d_k P_L \nu_i 
+\tilde d_{Rk}^\dagger \bar \nu_i^c P_L d_j 
\nonumber\\
&& \hspace{2.5em}
-\tilde e_{Li} \bar d_k P_L u_j 
- \tilde u_{Lj} \bar d_k P_L e_i 
- \tilde d_{Rk}^\dagger \bar e_i^c P_L u_j 
\Bigr] 
\nonumber\\
&& \hspace{12em}
+ ({\rm H.c.}) \Biggr\}
\ ,
\label{eq:rpvlagrangian}
\end{eqnarray}
where field operators with tilde denote the sparticles.
The coupling constants $\lambda_{ijk} , \lambda'_{ijk} $ ($i,j,k = 1,2,3$ are the flavor indices) may have nonzero CP phases.
In this case, a nonzero CP-odd $e$-$N$ interaction is generated through the $t$-channel process [see Fig. \ref{fig:diagrams} (a)].
The matching with the CP-odd electron-quark couplings is given by
\begin{eqnarray}
-\frac{G_F}{\sqrt{2}}C_j^{\rm SP}
&=&
\sum_{i=2,3} \frac{{\rm Im} (\lambda'_{ijj} \lambda^*_{i11})}{2m_{\tilde \nu_i}^2 }
,
\\
-\frac{G_F}{\sqrt{2}}C_j^{\rm PS}
&=&
-\sum_{i=2,3} \frac{{\rm Im} (\lambda'_{ijj} \lambda^*_{i11})}{2m_{\tilde \nu_i}^2 }
,
\end{eqnarray}
where $j=d,s,b$.
We note that the RPV interactions do not contribute to the CP-odd electron-up-type quark interaction at the tree level.
The PS type CP-odd electron-$d$-quark interaction is enhanced by the pion pole effect of the nucleon pseudoscalar density $\langle N|\bar d i \gamma_5 d|N\rangle$ at the nucleon level.
The constraints given from the experimental data of $d_{\rm Hg}$ are then
\begin{eqnarray}
\Biggl| \sum_{i=2,3} \frac{{\rm Im} (\lambda'_{i11} \lambda^*_{i11})}{m_{\tilde \nu_i}^2 } \Biggr| 
&<&
1.2 \times 10^{-8}\, {\rm TeV}^{-2}
,
\label{eq:rpvlimit1}
\\
\Biggl| \sum_{i=2,3} \frac{{\rm Im} (\lambda'_{i22} \lambda^*_{i11})}{m_{\tilde \nu_i}^2 } \Biggr| 
&<&
1.6 \times 10^{-6}\, {\rm TeV}^{-2}
,
\\
\Biggl| \sum_{i=2,3} \frac{{\rm Im} (\lambda'_{i33} \lambda^*_{i11})}{m_{\tilde \nu_i}^2 } \Biggr| 
&<&
1.7 \times 10^{-4}\, {\rm TeV}^{-2}
.
\end{eqnarray}
The flavor hierarchy of the above upper limits is due to the nucleon matrix elements which become smaller for heavier flavor.
From the LHC experimental data, the mass of sneutrinos are constrained by $m_{\tilde \nu_i} > 1$ TeV \cite{Aaboud:2018zeb}.
The coupling $\sum_{i=2,3} \frac{{\rm Im} (\lambda'_{i11} \lambda^*_{i11})}{m_{\tilde \nu_i}^2 }$ [Eq. (\ref{eq:rpvlimit1})] can be specifically constrained by the EDM of $^{199}$Hg via the CP-odd $e$-$N$ interaction thanks to the pion pole enhancement.
On the other hand, heavier flavor contributions are more strongly constrained by the analysis of the Barr-Zee type diagrams \cite{rpvbarr-zee1,rpvbarr-zee2,yamanakabook,rpvlinearprogramming}.


The analysis of the CP-odd $e$-$N$ interaction also has an important impact on the CP violation of the leptoquark models \cite{leptoquark1,leptoquark2,leptoquark3,leptoquark4}.
It is generated by the $s$-channel process, as shown in Fig. \ref{fig:diagrams} (b).
We can conceive spin-1 and spin-0 leptoquarks, as given by the Lagrangian of Ref. \cite{herczegleptoquark}:
\begin{eqnarray}
{\cal L }_{\rm LQ} 
&=&
S_1
[
g'_{1L} (\bar u^c P_L e  - \bar d^c P_L \nu_e )
+g'_{1R} \bar u^c P_R e
]
\nonumber\\
&& 
-(R_2)_+ 
[
h'_{2L} \bar u P_L e 
+h'_{2R} 
\bar u P_R e 
]
\nonumber\\
&& +
U_1^\mu 
[
h'_{1L} (\bar u \gamma_\mu P_L \nu_e + \bar d \gamma_\mu P_L e)
+h'_{1R} \bar d \gamma_\mu P_R e 
]
\nonumber\\
&& 
-(V_2)_+^\mu 
[
g'_{2L} \bar d^c \gamma_\mu P_L e
+g'_{2R}\bar d^c \gamma_\mu P_R e 
]
\nonumber\\
&&
+ ({\rm H.c.}) 
\ ,
\label{eq:lqlagrangian}
\end{eqnarray}
where ${S_1}$ and ${R_2}$ (${U_1}$ and ${V_2}$) are the spin-0 (spin-1) leptoquarks.
The spin-1 leptoquarks can only generate SP and PS type CP-odd $e$-$N$ interactions, whereas the spin-0 one also gives a T type one.
The matching with the CP-odd electron-quark couplings is given by
\begin{eqnarray}
-\frac{G_F}{\sqrt{2}}C_u^{\rm SP}
&=&
-\frac{{\rm Im} (h'_{2L} {h'}^*_{2R})}{4m_{R_1}^2 }
-\frac{{\rm Im} (g'_{1L} {g'}^*_{1R})}{4m_{S_1}^2 }
,
\\
-\frac{G_F}{\sqrt{2}}C_u^{\rm PS}
&=&
-\frac{{\rm Im} (h'_{2L} {h'}^*_{2R})}{4m_{R_1}^2 }
-\frac{{\rm Im} (g'_{1L} {g'}^*_{1R})}{4m_{S_1}^2 }
,
\\
\frac{G_F}{2\sqrt{2}}C_u^{\rm T}
&=&
-\frac{{\rm Im} (h'_{2L} {h'}^*_{2R})}{8m_{R_1}^2 }
+\frac{{\rm Im} (g'_{1L} {g'}^*_{1R})}{8m_{S_1}^2 }
,
\\
-\frac{G_F}{\sqrt{2}}C_d^{\rm SP}
&=&
\frac{{\rm Im} (h'_{1L} {h'}^*_{1R})}{m_{U_1}^2 }
+\frac{{\rm Im} (g'_{2L} {g'}^*_{2R})}{m_{V_1}^2 }
,
\\
-\frac{G_F}{\sqrt{2}}C_d^{\rm PS}
&=&
-\frac{{\rm Im} (h'_{1L} {h'}^*_{1R})}{m_{U_1}^2 }
-\frac{{\rm Im} (g'_{2L} {g'}^*_{2R})}{m_{V_1}^2 }
.
\end{eqnarray}
We can then derive the following constraints from the experimental data of $d_{\rm Hg}$:
\begin{eqnarray}
\Biggl| \frac{{\rm Im} ( g'_{1L} {g'}^*_{1R})}{m_{S_1}^2 } \Biggr| 
&<&
2.8 \times 10^{-8}\, {\rm TeV}^{-2}
,
\\
\Biggl| \frac{{\rm Im} ( h'_{2L} {h'}^*_{2R})}{m_{R_2}^2 } \Biggr| 
&<&
2.5 \times 10^{-8}\, {\rm TeV}^{-2}
,
\\
\Biggl| \frac{{\rm Im} ( h'_{1L} {h'}^*_{1R})}{m_{U_1}^2 } \Biggr| 
&<&
6.2 \times 10^{-9}\, {\rm TeV}^{-2}
,
\\
\Biggl| \frac{{\rm Im} ( g'_{2L} {g'}^*_{2R})}{m_{V_1}^2 } \Biggr| 
&<&
6.2 \times 10^{-9}\, {\rm TeV}^{-2}
,
\end{eqnarray}
where $m_{S_1}$ and $m_{R_1}$ ($m_{U_1}$ and $m_{V_1}$) are the masses of the spin-0 (spin-1) leptoquarks. 
The masses of the leptoquarks are constrained to be larger than 1 TeV by the null result of LHC experiments \cite{Aaboud:2016qeg,Sirunyan:2018ruf,Sirunyan:2018ryt,Sirunyan:2018btu}.
The above leptoquark interactions also contribute to the electron EDM \cite{Barr:1987sp,Geng:1990gr,fuyuto}, but the upper limits are looser by several orders even if we consider the latest experimental constraint $d_e < 1.1 \times 10^{-29}e$ cm \cite{Andreev:2018ayy}, if the fermions are restricted to the first generation.
Here for simplicity, we only considered one leptoquark multiplet for each spin.
In the general case with several leptoquark multiplets, there may be additional mixings \cite{herczegleptoquark}.
For the exchange of spin-1 leptoquarks, the EDM of $^{199}$Hg is dominantly generated by the PS type interaction, due to the pion pole enhancement.
In the case of spin-0 leptoquark exchange, the contributions from the T and PS type CP-odd $e$-$N$ interactions are comparable.


Let us also mention the SM contribution, generated by the CP phase of the CKM matrix.
The current understanding is that its leading effect to the CP-odd $e$-$N$ interaction is given by the pion-exchange process, via the SP type interaction
($C_N^{\rm SP}$) \cite{hemackellarpakvasa,pospelovsmatomicedm,yamanakasmedm}.
The coupling is estimated to be 
\begin{equation}
C_N^{\rm SP} \sim 10^{-17}
,
\end{equation}
which gives a contribution of $d_{\rm Hg} \sim 10^{-39}e$ cm.
In the diamagnetic atomic system, the contribution from the nuclear Schiff moment \cite{schiff} is also relevant.
From the most recent calculation, it is given by $d_{\rm Hg} \sim 10^{-35}e$ cm \cite{yamanakasmedm}, which is much larger than that of $C_N^{\rm SP}$.
However, we have to note that the above analyses did not consider the dynamical nuclear effect of the intermediate hypernuclear state \cite{yamanakasmedm2}, and the neglect of it may introduce sizable systematics.
This issue has to be inspected in future works.

\section{\label{sec:conclusion}Conclusion}

In summary, we evaluated the nuclear spin matrix elements of $^{199}$Hg, which are required in the evaluation of the atomic EDM generated by the PS and T type CP-odd $e$-$N$ interactions within the nuclear shell model.
The $^{199}$Hg nucleus has a dominant configuration with $p_{1/2}$ orbital neutron, yielding $\langle \Psi |\, \sigma_{nz} | \Psi \rangle \approx -0.4$. 
For the proton one, the errorbar is larger than the central value, but its size is much smaller than the neutron one, so it can be neglected in the analysis of many candidates of new physics beyond the SM.
Our result is consistent with the simple shell model picture with only one valence neutron, but we emphasize that this was obtained with the large-scale shell-model that considers the dynamical effects among nucleons in the core, reducing the systematics and greatly increasing its reliability.
Before this work, the nuclear spin matrix elements of $^{199}$Hg were the only missing link between the semi-leptonic CP violation and the experimental data of the EDM of $^{199}$Hg.
Through our analysis, we could fill the gap, and combining with the accurate hadronic and atomic level inputs, the CP-odd $e$-$N$ interaction became the most accurately known CP-odd process of the EDM of $^{199}$Hg.

We also analyzed the contribution of the P, CP-odd $e$-$N$ interaction to the EDM of the $^{199}$Hg atom within 2HDMs, supersymmetric models, and leptoquark models. 
Thanks to the tight experimental limit of the $^{199}$Hg atom, we could set strong constraints on the semi-leptonic sector of these models. 
The CP-odd $e$-$N$ interaction is therefore critically important to probe those specific candidates.

Our analysis using the nuclear shell model definitely has to be extended to the study of the nuclear Schiff moment generated by the intrinsic nucleon EDM and the CP-odd pion-nucleon interactions to reduce the theoretical uncertainty.
We also expect our framework to be applicable in the evaluation of the spin matrix elements of other heavy nuclei such as $^{225}$Ra which has already been measured in experiment \cite{bishof}.

\begin{acknowledgments}
The authors thank Takeshi Inoue for useful discussions.
This work was supported by Grant-in-Aid for Scientific Research (C) (Grants No. 16K05341 and No. 17K05450) from the Japan Society for the Promotion of Science (JSPS).
This is also supported by the JSPS Postdoctoral Fellowships for Research Abroad.
\end{acknowledgments}


\begin{thebibliography}{99}

\bibitem{sakharov}
A. D. Sakharov, Pisma Zh. Eksp. Teor. Fiz. {\bf 5}, 32 (1967) [JETP Lett. {\bf 5}, 24 (1967)].

\bibitem{farrar}
G. R. Farrar and M. E. Shaposhnikov, Phys. Rev. D {\bf 50}, 774 (1994).

\bibitem{huet}
P. Huet and E. Sather, Phys. Rev. D {\bf 51}, 379 (1995).

\bibitem{hereview}
X.-G. He, B. H. J. McKellar, and S. Pakvasa, Int. J. Mod. Phys. A {\bf 4}, 5011 (1989) [Erratum ibid. A {\bf 6}, 1063 (1991)].

\bibitem{bernreutherreview}
W. Bernreuther and M. Suzuki, Rev. Mod. Phys. {\bf 63}, 313 (1991).

\bibitem{khriplovichbook}
I. B. Khriplovich and S. K. Lamoreaux, {\it CP Vioaltion Without Strangeness}, Springer, Berlin
(1997).

\bibitem{ginges}
J. S. M. Ginges and V. V. Flambaum, Phys. Rep. {\bf 397}, 63 (2004).

\bibitem{pospelovreview}
M. Pospelov and A. Ritz, Ann. Phys. (Amsterdam) {\bf 318}, 119 (2005).

\bibitem{raidal}
M. Raidal {\it et al.}, Eur. Phys. J. C {\bf 57}, 13 (2008).

\bibitem{naviliatreview}
O. Naviliat-Cuncic and R. G. E. Timmermans, Comptes Rendus Physique {\bf 13}, 168 (2012).

\bibitem{fukuyama}
T. Fukuyama, Int. J. Mod. Phys. A {\bf 27}, 1230015 (2012).

\bibitem{engeledmreview}
J. Engel, M. J. Ramsey-Musolf, and U. van Kolck, Prog. Part. Nucl. Phys. {\bf 71}, 21 (2013).

\bibitem{jungmann}
K. Jungmann, Ann. Phys. (Berlin) {\bf 525}, 550 (2013).

\bibitem{yamanakabook}
N.~Yamanaka,
  ``Analysis of the Electric Dipole Moment in the R-parity Violating Supersymmetric Standard Model,''
Springer, Berlin Germany (2014).
  doi:10.1007/978-4-431-54544-6

\bibitem{roberts}
B. M. Roberts, V. A. Dzuba, and V. V. Flambaum, Annu. Rev. Nucl. Part. Sci. {\bf 65}, 63 (2015).

\bibitem{yamanakanuclearedmreview}
N. Yamanaka, Int. J. Mod. Phys. E {\bf 26}, 1730002 (2017).

\bibitem{atomicedmreview}
N. Yamanaka, B. Sahoo, N. Yoshinaga, T. Sato, K. Asahi, and B. Das, Eur. Phys. J. A {\bf 53}, 54 (2017).

\bibitem{chuppreview}
T. E. Chupp, P. Fierlinger, M. J. Ramsey-Musolf, and J. T. Singh, Rev. Mod. Phys. {\bf 91}, 015001 (2019).

\bibitem{safronova}
M. S. Safronova, D. Budker, D. DeMille, D. F. Jackson Kimball, A. Derevianko, and C. W. Clark, Rev. Mod. Phys. {\bf 90}, 025008 (2018).

\bibitem{orzel}
C. Orzel, arXiv:1802.06334 [physics.atom-ph].

\bibitem{ckm}
M. Kobayashi and T. Maskawa, Prog. Theor. Phys. {\bf 49}, 652 (1973).

\bibitem{jarlskog}
C. Jarlskog, Phys. Rev. Lett. {\bf 55}, 1039 (1985).

\bibitem{regan}
B. C. Regan, E. D. Commins, C. J. Schmidt, and D. DeMille, Phys. Rev. Lett. {\bf 88}, 071805 (2002).

\bibitem{baker}
C. A. Baker {\it et al.}, Phys. Rev. Lett. {\bf 97}, 131801 (2006).

\bibitem{muong2}
G. W. Bennett {\it et al.} (Muon $g-2$ Collaboration), Phys. Rev. D {\bf 80}, 052008 (2009).

\bibitem{hudson1}
J. J. Hudson, D. M. Kara, I. J. Smallman, B. E. Sauer, M. R. Tarbutt, and E. A. Hinds, Nature (London) {\bf 473}, 493 (2011). 

\bibitem{hudson2}
J. J. Hudson, D. M. Kara, I. J. Smallman, B. E. Sauer, M. R. Tarbutt, and E. A. Hinds, New J. Phys. {\bf 14}, 103051 (2012).

\bibitem{acme}
J. Baron {\it et al.} (ACME Collaboration), Science {\bf 343}, 269 (2014). 

\bibitem{Andreev:2018ayy} 
V.~Andreev {\it et al.} (ACME Collaboration), Nature {\bf 562}, 355 (2018).

\bibitem{chin}
C. Chin, V. Leiber, V. Vuleti\'{c}, A. J. Kerman, and S. Chu, Phys. Rev. A {\bf 63}, 033401 (2001).

\bibitem{sakemi}
Y. Sakemi {\it et al.}, J. Phys. Conf. Ser. {\bf 302}, 012051 (2011).

\bibitem{storage1}
I. B. Khriplovich, Phys. Lett. B {\bf 444}, 98 (1998).

\bibitem{storage2}
F. J. M. Farley {\it et al}., Phys. Rev. Lett. {\bf 93}, 052001 (2004).

\bibitem{storage4}
Y. F. Orlov, W. M. Morse, and Y. K. Semertzidis, Phys. Rev. Lett. {\bf 96}, 214802 (2006).

\bibitem{Anastassopoulos}
V. Anastassopoulos {\it et al.}, Rev. Sci. Instrum. {\bf 87}, 115116 (2016). 

\bibitem{jedi2}
N. Hempelmann {\it et al.} (JEDI Collaboration), Phys. Rev. Lett. {\bf 119}, 014801 (2017).

\bibitem{botella}
F. J. Botella, L. M. Garcia Martin, D. Marangotto, F. M. Vidal, A. Merli, N. Neri, A. Oyanguren, and J. R. Vidal, Eur. Phys. J. C {\bf 77}, 181 (2017). 

\bibitem{Baryshevsky}
V. G. Baryshevsky, arXiv:1803.05770 [hep-ph].

\bibitem{xinchen}
X. Chen and Y. Wu, arXiv:1803.00501 [hep-ph].

\bibitem{koksal}
M. K\"{o}ksal, A. A. Billur, A. Guti\'{e}rrez-Rodr\'{i}guez, and M. A. Hern\'{a}ndez-Ru\'{i}z, Phys. Rev. D {\bf 98}, 015017 (2018).

\bibitem{inertgasmatrix}
A. C. Vutha, M. Horbatsch, and E. A. Hessels, Atoms {\bf 6}, 3 (2018).

\bibitem{Kozyryev:2017cwq} 
I.~Kozyryev and N.~R.~Hutzler, Phys.\ Rev.\ Lett.\  {\bf 119}, 133002 (2017).

\bibitem{Sachdeva}
N. Sachdeva {\it et al.}, arXiv:1902.02864 [physics.atom-ph].

\bibitem{bishof}
M. Bishof {\it et al.}, Phys. Rev. C {\bf 94}, 025501 (2016). 

\bibitem{graner}
B. Graner, Y. Chen, E. G. Lindahl, and B. R. Heckel, Phys. Rev. Lett. {\bf 116}, 161601 (2016) [Erratum ibid. {\bf 119}, 119901 (2017)]. 

\bibitem{Araujo:2015zsa} 
J.~B.~Araujo, R.~Casana, and M.~M.~Ferreira, Phys.\ Rev.\ D {\bf 92}, 025049 (2015).

\bibitem{Araujo:2016hsn} 
J.~B.~Araujo, R.~Casana, and M.~M.~Ferreira, Phys.\ Lett.\ B {\bf 760}, 302 (2016).

\bibitem{araujo}
J. B. Araujo, R. Casana, and M. M. Ferreira Jr., Phys. Rev. D {\bf 97}, 055032 (2018).

\bibitem{Mantry:2014zsa} 
S.~Mantry, M.~Pitschmann, and M.~J.~Ramsey-Musolf, Phys.\ Rev.\ D {\bf 90}, 054016 (2014).

\bibitem{stadnik}
Y. V. Stadnik, V. A. Dzuba, and V. V. Flambaum, Phys. Rev. Lett. {\bf 120}, 013202 (2018).

\bibitem{schiff}
L. I. Schiff, Phys. Rev. {\bf 132}, 2194 (1963).

\bibitem{bouchiat}
C. Bouchiat, Phys. Lett. B {\bf 57}, 284 (1975).

\bibitem{hinds}
E. A. Hinds, C. E. Loving, and G. P. H. Sandars, Phys. Lett. B {\bf 62}, 97 (1976).

\bibitem{martensson}
A. M. Martensson-Pendrill, Phys. Rev. Lett. {\bf 54}, 1153 (1985).

\bibitem{flambaumshellmodel2}
V. V. Flambaum and I. B. Khriplovich, Zh. Eksp. Teor. Fiz. {\bf 89}, 1505 (1985) [Sov. Phys. JETP {\bf 62}, 872 (1985)].

\bibitem{dzuba2}
V. A. Dzuba, V. V. Flambaum, and P. G. Silvestrov, Phys. Lett. B {\bf 154}, 93 (1985).

\bibitem{barr2}
S. M. Barr, Phys. Rev. D {\bf 45}, 4148 (1992).

\bibitem{hemckellar}
X.-G. He and B. McKellar, Phys. Lett. B {\bf 390}, 318 (1997).

\bibitem{barr3}
S. M. Barr, Phys. Rev. Lett. {\bf 68}, 1822 (1992).

\bibitem{hemackellarpakvasa}
X.-G. He, B. H. J. McKellar, and S. Pakvasa, Phys. Lett. B {\bf 283}, 348 (1992).

\bibitem{barr4}
S. M. Barr, Phys. Rev. D {\bf 47}, 2025 (1993).

\bibitem{jung1}
M. Jung and A. Pich, J. High Energy Phys. {\bf 1404} (2014) 076. 

\bibitem{fischler}
W. Fischler, S. Paban, and S. D. Thomas, Phys. Lett. B {\bf 289}, 373 (1992).

\bibitem{Lebedev:2002ne} 
O.~Lebedev and M.~Pospelov, Phys.\ Rev.\ Lett.\  {\bf 89}, 101801 (2002).

\bibitem{Demir:2003js} 
D.~A.~Demir, O.~Lebedev, K.~A.~Olive, M.~Pospelov, and A.~Ritz, Nucl.\ Phys.\ B {\bf 680}, 339 (2004).

\bibitem{pilaftsisgluoniccpve-n}
A. Pilaftsis, Nucl. Phys. B {\bf 644}, 263 (2002).

\bibitem{edmmssmreloaded}
J. Ellis, J. S. Lee, and A. Pilaftsis, J. High Energy Phys. {\bf 10}, 049 (2008).

\bibitem{Lee:2012wa}
J.~S.~Lee, M.~Carena, J.~Ellis, A.~Pilaftsis, and C.~E.~M.~Wagner, Comput.\ Phys.\ Commun.\  {\bf 184}, 1220 (2013).

\bibitem{herczege-n1}
P. Herczeg, Phys. Rev. D {\bf 61}, 095010 (2000).

\bibitem{herczege-n2}
J. Res. Natl. Inst. Stand. Technol. {\bf 110}, 453 (2005).

\bibitem{cpve-n1loop}
N. Yamanaka, Phys. Rev. D {\bf 85}, 115012 (2012).

\bibitem{rpvlinearprogramming}
N. Yamanaka, T. Sato, and T. Kubota, J. High Energy Phys. {\bf 1412} (2014) 110.

\bibitem{mahanta}
U. Mahanta, Phys. Lett. B {\bf 337}, 128 (1994).

\bibitem{herczegleptoquark}
P. Herczeg, Phys. Rev. D {\bf 68}, 116004 (2003) [Erratum ibid. {\bf 69}, 039901 (2004)].

\bibitem{fuyuto}
K. Fuyuto, M. Ramsey-Musolf, and T. Shen, Phys. Lett. B {\bf 788}, 52 (2019).

\bibitem{chiqcdsigmaterm}
Y.-B. Yang, A. Alexandru, T. Draper, J. Liang, and K.-F. Liu ($\chi$QCD Collaboration), Phys. Rev. D {\bf 94}, 054503 (2016).

\bibitem{rqcdsigmaterm}
G. S. Bali, S. Collins, D. Richtmann, A. Sch\"{a}fer, W. S\"{o}ldner, and A. Sternbeck (RQCD Collaboration), Phys. Rev. D {\bf 93}, 094504 (2016).

\bibitem{bmwsigmaterm}
S. Durr {\it et al.} (Budapest-Marseille-Wuppertal Collaboration), Phys. Rev. Lett. {\bf 116}, 172001 (2016).

\bibitem{etmsigmaterm}
A. Abdel-Rehim, C. Alexandrou, M. Constantinou, K. Hadjiyiannakou, K. Jansen, Ch. Kallidonis, G. Koutsou, and A. Vaquero Avil\'{e}s-Casco (ETM Collaboration), Phys. Rev. Lett. {\bf 116}, 252001 (2016).

\bibitem{green}
J. R. Green, J. W. Negele, A. V. Pochinsky, S. N. Syritsyn, M. Engelhardt, and S. Krieg, Phys. Rev. D {\bf 86}, 114509 (2012).

\bibitem{rqcdisovector}
G. S. Bali, S. Collins, B. Gl\"{a}{\ss}le, M. G\"{o}ckeler, J. Najjar, R. H. R\"{o}dl, A. Sch\"{a}fer, R. W. Schiel, W. S\"{o}ldner, and A. Sternbeck (RQCD Collaboration), Phys. Rev. D {\bf 91}, 054501 (2015).

\bibitem{Yamanaka:2015lfk}
N. Yamanaka, H. Ohki, S. Hashimoto, and T. Kaneko (JLQCD Collaboration),
PoS LATTICE {\bf 2015} (2016) 121
[arXiv:1511.04589 [hep-lat]].

\bibitem{chiqcdisovector}
Y.-B. Yang, A. Alexandru, T. Draper, M. Gong, and K.-F. Liu ($\chi$QCD Collaboration), Phys. Rev. D {\bf 93}, 034503 (2016).

\bibitem{pndmeisovector}
T. Bhattacharya, V. Cirigliano, S. D. Cohen, R. Gupta, H.-W. Lin, and B. Yoon (PNDME Collaboration), Phys. Rev. D {\bf 94}, 054508 (2016).

\bibitem{pndmetensor}
R. Gupta, B. Yoon, T. Bhattacharya, V. Cirigliano, Y.-C. Jang, and H.-W. Lin (PNDME Collaboration), Phys. Rev. D {\bf 98}, 091501 (2018).

\bibitem{etm2017}
C. Alexandrou {\it et al.}, Phys. Rev. D {\bf 95}, 114514 (2017) [Erratum ibid. {\bf 96}, 099906 (2017)].

\bibitem{dzuba}
V. A. Dzuba, V. V. Flambaum, and S. G. Porsev, Phys. Rev. A {\bf 80}, 032120 (2009).

\bibitem{latha}
K. V. P. Latha, D. Angom, B. P. Das, and D. Mukherjee, Phys. Rev. Lett. {\bf 103}, 083001 (2009) [Erratum ibid. {\bf 115}, 059902 (2015)].

\bibitem{singh}
Y. Singh and B. K. Sahoo, Phys. Rev. A {\bf 91}, 030501 (2015). 

\bibitem{radziute}
L. Radziute, G. Gaigalas, P. J\"{o}nsson, and J. Bieron, Phys. Rev. A {\bf 93}, 062508 (2016).

\bibitem{sahoo}
B. K. Sahoo, Phys. Rev. D {\bf 95}, 013002 (2017).

\bibitem{sahoo2}
B. K. Sahoo and B. P. Das, Phys. Rev. Lett. {\bf 120}, 203001 (2018).

\bibitem{flambaumshellmodel1}
O. P. Sushkov, V. V. Flambaum, I. B. Khriplovich, Zh. Eksp. Teor. Fiz. {\bf 87}, 1521 (1984) [Sov. Phys. JETP {\bf 60}, 873 (1984)].

\bibitem{flambaumshellmodel3}
V. V. Flambaum, I. B. Khriplovich, and O. P. Sushkov, Nucl. Phys. A {\bf 449}, 750 (1986).

\bibitem{yoshinaga1}
N. Yoshinaga, K. Higashiyama, and R. Arai, Prog. Theor. Phys. {\bf 124}, 1115 (2010).

\bibitem{yoshinaga2}
N. Yoshinaga, K. Higashiyama, R. Arai, and E. Teruya, Phys. Rev. C {\bf 87}, 044332 (2013) [Erratum ibid. {\bf 89}, 069902 (2014)]. 

\bibitem{yoshinaga3}
N. Yoshinaga, K. Higashiyama, R. Arai, and E. Teruya, Phys. Rev. C {\bf 89}, 045501 (2014).

\bibitem{teruya}
E. Teruya, N. Yoshinaga, K. Higashiyama, and K. Asahi, Phys. Rev. C {\bf 96}, 015501 (2017). 

\bibitem{lathaxerayb}
K. V. P. Latha and P. R. Amjith, Phys. Rev. A {\bf 87}, 022509 (2013).

\bibitem{fleig}
T. Fleig and M. Jung, J. High Energy Phys. {\bf 1807} (2018) 012.

\bibitem{bacchetta}
A. Bacchetta, A. Courtoy, and M. Radici, J. High Energy Phys. {\bf 1303} (2013) 119.

\bibitem{radici}
M. Radici, A. Courtoy, A. Bacchetta, and M. Guagnelli, J. High Energy Phys. {\bf 1505} (2015) 123. 

\bibitem{kang}
Z.-B. Kang, A. Prokudin, P. Sun, and F. Yuan, Phys. Rev. D {\bf 93}, 014009 (2016).

\bibitem{radici2}
M. Radici and A. Bacchetta, Phys. Rev. Lett. {\bf 120}, 192001 (2018).

\bibitem{courtoy}
A. Courtoy, S. Bae{\ss}ler, M. Gonz\'{a}lez-Alonso, and S. Liuti, Phys. Rev. Lett. {\bf 115}, 162001 (2015).

\bibitem{yez}
Z. Ye, N. Sato, K. Allada, T. Liu, J.-P. Chen, H. Gao, Z.-B. Kang, A. Prokudin, P. Sun, and F. Yuan, Phys. Lett. B {\bf 767}, 91 (2017). 

\bibitem{accardi}
A. Accardi and A. Bacchetta, Phys. Lett. B {\bf 773}, 632 (2017). 

\bibitem{gaotensor}
T. Liu, Z. Zhao, and H. Gao, Phys. Rev. D {\bf 97}, 074018 (2018). 

\bibitem{barone}
V. Barone, Phys. Lett. B {\bf 409}, 499 (1997).

\bibitem{artru}
X. Artru and M. Mekhfi, Z. Phys. C {\bf 45}, 669 (1990).

\bibitem{degrassi}
G. Degrassi, E. Franco, S. Marchetti, and L. Silvestrini, J. High Energy Phys. {\bf 11}, 044 (2005).

\bibitem{renormalizationedm}
J. Hisano, K. Tsumura, and M. J. S. Yang, Phys. Lett. B {\bf 713}, 473 (2012).

\bibitem{renormalizationedm2}
W. Dekens and J. de Vries, J. High Energy Phys. {\bf 1305}, 149 (2013).

\bibitem{tensorsde}
N. Yamanaka, T. M. Doi, S. Imai, and H. Suganuma, Phys. Rev. D {\bf 88}, 074036 (2013).

\bibitem{pitschmann}
M. Pitschmann, C.-Y. Seng, C. D. Roberts, and S. M. Schmidt, Phys. Rev. D {\bf 91}, 074004 (2015).

\bibitem{alonso}
M. Gonz\'{a}lez-Alonso and J. Martin Camalich, Phys. Rev. Lett. {\bf 112}, 042501 (2014).

\bibitem{cheng1}
H.-Y. Cheng, Phys. Lett. B {\bf 219}, 347 (1989).

\bibitem{chengli}
T. P. Cheng and L. F. Li, Phys. Rev. Lett. {\bf 62}, 1441 (1989).

\bibitem{dienes}
K. R. Dienes, J. Kumar, B. Thomas, and D. Yaylali, Phys. Rev. D {\bf 90}, 015012 (2014).

\bibitem{cheng2}
H.-Y. Cheng and C.-W. Chiang, J. High Energy Phys. {\bf 1207} (2012) 009.

\bibitem{scopel}
S. Scopel and H. Yu, JCAP {\bf 1704} (2017) 031. 

\bibitem{pdg}
M. Tanabashi {\it et al.} (Particle Data Group), Phys. Rev. D {\bf 98}, 030001 (2018).

\bibitem{fermilablattice}
A. Bazavov {\it et al.} (Fermilab Lattice, MILC, and TUMQCD Collaborations), Phys. Rev. D {\bf 98}, 054517 (2018).

\bibitem{Campos:2018ahf} 
I.~Campos {\it et al.} (ALPHA Collaboration), Eur.\ Phys.\ J.\ C {\bf 78}, 387 (2018).

\bibitem{axialsde}
N. Yamanaka, S. Imai, T. M. Doi, and H. Suganuma, Phys. Rev. D {\bf 89}, 074017 (2014).

\bibitem{witten}
E. Witten, Nucl. Phys. B {\bf 104}, 445 (1976).

\bibitem{shifman}
M. A. Shifman, A. I. Vainshtein, and V. I. Zakharov, Phys. Lett. B {\bf 78}, 443 (1978).

\bibitem{hatsuda1}
T. Hatsuda and T. Kunihiro, Nucl. Phys. B {\bf 387}, 715 (1992).

\bibitem{hatsuda2}
T. Hatsuda and T. Kunihiro, Phys. Rep. {\bf 247}, 221 (1994).

\bibitem{Zhitnitsky}
A. R. Zhitnitsky, Phys. Rev. D {\bf 55}, 3006 (1997).

\bibitem{alarcon1}
J. M. Alarc\'{o}n, J. Martin Camalich, and J. A. Oller, Phys. Rev. D {\bf 85}, 051503 (2012).

\bibitem{alarcon2}
J. M. Alarc\'{o}n, J. Martin Camalich, and J. A. Oller, Ann. Phys. (Amsterdam) {\bf 336}, 413 (2013).

\bibitem{Hoferichter}
M. Hoferichter, J. Ruiz de Elvira, B. Kubis, and U.-G. Mei{\ss}ner, Phys. Rev. Lett. {\bf 115}, 092301 (2015). 

\bibitem{Hoferichter2}
M. Hoferichter, J. Ruiz de Elvira, B. Kubis, and U.-G. Mei{\ss}ner, Phys. Lett. B {\bf 760}, 74 (2016).

\bibitem{yao}
D.-L. Yao, D. Siemens, V. Bernard, E. Epelbaum, A. M. Gasparyan, J. Gegelia, H. Krebs, and U.-G. Mei{\ss}ner, J. High Energy Phys. {\bf 1605} (2016) 038.

\bibitem{deelvira}
J. Ruiz de Elvira, M. Hoferichter, B. Kubis, and U.-G. Mei{\ss}ner, J. Phys. G {\bf 45}, 024001 (2018).

\bibitem{ohki1}
H. Ohki {\it et al.} (JLQCD Collaboration), Phys. Rev. D {\bf 78}, 054502 (2008).

\bibitem{ohki2}
K. Takeda, S. Aoki, S. Hashimoto, T. Kaneko, J. Noaki, and T. Onogi (JLQCD Collaboration), Phys. Rev. D {\bf 83}, 114506 (2011).

\bibitem{ohki3}
H. Ohki, K. Takeda, S. Aoki, S. Hashimoto, T. Kaneko, H. Matsufuru, J. Noaki, and T. Onogi (JLQCD Collaboration), Phys. Rev. D {\bf 87}, 034509 (2013).

\bibitem{alarcon3}
J. M. Alarc\'{o}n, L. S. Geng, J. Martin Camalich, and J. A. Oller, Phys. Lett. B {\bf 730}, 342 (2014).

\bibitem{lattice_charm_content1}
W. Freeman and D. Toussaint (MILC Collaboration), Phys. Rev. D {\bf 88}, 054503 (2013).

\bibitem{lattice_charm_content2}
M. Gong {\it et al.} ($\chi$QCD Collaboration), Phys. Rev. D {\bf 88}, 014503 (2013).

\bibitem{hobbs}
T. J. Hobbs, M. Alberg, and G. A. Miller, Phys. Rev. D {\bf 96}, 074023 (2017).

\bibitem{Higashi03} 
K. Higashiyama, N. Yoshinaga, and K. Tanabe, Phys. Rev. C {\bf 67}, 044305 (2003).

\bibitem{Yoshi04} 
N. Yoshinaga and K. Higashiyama, Phys. Rev. C {\bf 69}, 054309 (2004).

\bibitem{Higashi11-1}
K. Higashiyama and N. Yoshinaga, Phys. Rev. C {\bf 83}, 034321 (2011).

\bibitem{Higashi11-2}
K. Higashiyama and N. Yoshinaga, Phys. Rev. C {\bf 89}, 049903 (2014).

\bibitem{ENSDF}
``Evaluated Nuclear Structure Data File (ENSDF)''
\url{www.nndc.bnl.gov/ensdf/}

\bibitem{NDS205}
F. G. Kondev, Nuclear Data Sheets {\bf 101}, 521 (2004).

\bibitem{NDS203}
F. G. Kondev, Nuclear Data Sheets {\bf 105}, 1 (2005).

\bibitem{NDS201}
F. G. Kondev, Nuclear Data Sheets {\bf 108}, 365 (2007).

\bibitem{NDS199}
B. Singh, Nuclear Data Sheets {\bf 108}, 79 (2007).

\bibitem{teruya2}
E. Teruya, K. Higashiyama, and N. Yoshinaga, Phys. Rev. C {\bf 93}, 064327 (2016).

\bibitem{yanase}
K. Yanase, E. Teruya, K. Higashiyama, and N. Yoshinaga, Phys. Rev. C {\bf 98}, 014308 (2018).

\bibitem{towner}
I. S. Towner, Phys. Rep. {\bf 155}, 263 (1987).

\bibitem{chemtob}
M. Chemtob, Nucl. Phys. A {\bf 123}, 449 (1969).

\bibitem{barzakh}
A. E. Barzakh, D. V. Fedorov, V. S. Ivanov, P. L. Molkanov, F. V. Moroz, S. Yu. Orlov, V. N. Panteleev, M. D. Seliverstov, and Yu. M. Volkov, Phys. Rev. C {\bf 97}, 014322 (2018).

\bibitem{weinberg}
S. Weinberg, Phys. Lett. B {\bf 251}, 288 (1990).

\bibitem{park}
T.-S. Park, D.-P. Min, and M. Rho, Nucl. Phys. A {\bf 596}, 515 (1996).

\bibitem{tdlee}
T. D. Lee, Phys. Rev. D {\bf 8}, 1226 (1973).

\bibitem{weinbergoperator}
S. Weinberg, Phys. Rev. Lett. {\bf 63}, 2333 (1989).

\bibitem{gunionreview}
J. F. Gunion, H. E. Haber, G. L. Kane, and S. Dawson, Front. Phys. {\bf 80}, 1 (2000).

\bibitem{branco}
G. C. Branco {\it et al.}, Phys. Rep. {\bf 516}, 1 (2012).

\bibitem{ivanov}
I. P. Ivanov, Prog. Part. Nucl. Phys. {\bf 95}, 160 (2017).

\bibitem{YongchaoZhang}
L. Bian, N. Chen, and Y. Zhang, Phys. Rev. D {\bf 96}, 095008 (2017).

\bibitem{ruisantos}
D. Fontes, M. M\"{u}hlleitner, J. C. Rom\~{a}o, R. Santos, J. P. Silva, and J. Wittbrodt, J. High Energy Phys. {\bf 1802} (2018) 073.

\bibitem{higgs1}
G. Aad {\it et al.} (ATLAS Collaboration), Phys. Lett. B {\bf 716}, 1 (2012).

\bibitem{higgs2}
S. Chatrchyan {\it et al.} (CMS Collaboration), Phys. Lett. B {\bf 716}, 30 (2012).

\bibitem{higgs3}
G. Aad {\it et al.} (ATLAS and CMS Collaborations), Phys. Rev. Lett. {\bf 114}, 191803 (2015).

\bibitem{Weinberg:1990me} 
S.~Weinberg, Phys.\ Rev.\ D {\bf 42}, 860 (1990).
  
\bibitem{Kao:1992jv} 
C.~Kao and R.~M.~Xu, Phys.\ Lett.\ B {\bf 296}, 435 (1992).

\bibitem{Aaboud:2017sjh} 
M.~Aaboud {\it et al.} (ATLAS Collaboration), J. High Energy Phys. {\bf 1801}, 055 (2018).

\bibitem{Sirunyan:2018zut} 
A.~M.~Sirunyan {\it et al.} (CMS Collaboration), J. High Energy Phys. {\bf 1809}, 007 (2018).
  
\bibitem{ellishwang}
J. Ellis, D. S. Hwang, K. Sakurai, and M. Takeuchi, J. High Energy Phys. {\bf 1404} (2014) 004.

\bibitem{demartin}
F. Demartin, F. Maltoni, K. Mawatari, B. Page, and M. Zaro, Eur. Phys. J. C {\bf 74}, 3065 (2014).

\bibitem{Chien:2015xha} 
Y.~T.~Chien, V.~Cirigliano, W.~Dekens, J.~de Vries, and E.~Mereghetti, JHEP {\bf 1602}, 011 (2016).

\bibitem{cirigliano}
V. Cirigliano, W. Dekens, J. de Vries, and E. Mereghetti, Phys. Rev. D {\bf 94}, 016002 (2016).

\bibitem{cirigliano2}
V. Cirigliano, W. Dekens, J. de Vries, and E. Mereghetti, Phys. Rev. D {\bf 94}, 034031 (2016).

\bibitem{mileo}
N. Mileo, K. Kiers, A. Szynkman, D. Crane, and E. Gegner, J. High Energy Phys. {\bf 1607} (2016) 056.

\bibitem{englert}
C. Englert, K. Nordstr\"{o}m, K. Sakurai, and M. Spannowsky, Phys. Rev. D {\bf 95}, 015018 (2017).

\bibitem{coleppa}
B. Coleppa, M. Kumar, S. Kumar, and B. Mellado, Phys. Lett. B {\bf 770}, 335 (2017).

\bibitem{Fuyuto:2017xup} 
K.~Fuyuto and M.~Ramsey-Musolf, Phys.\ Lett.\ B {\bf 781}, 492 (2018).

\bibitem{cpvhiggseleposi}
W. Bernreuther, L. Chen, I. Garc\'{i}a, M. Perell\'{o}, R. Poeschl F. Richard, E. Ros, and M. Vos, Eur. Phys. J. C {\bf 78}, 155 (2018).

\bibitem{Brandstetter:2018eju}
J. Brandstetter (ATLAS and CMS Collaborations), arXiv:1801.07926 [hep-ex].

\bibitem{goncalves}
D. Gon\c{c}alves, J. H. Kim, and K. Kong, J. High Energy Phys. {\bf 1806} (2018) 079.

\bibitem{barr-zee}
S. M. Barr and A. Zee, Phys. Rev. Lett. {\bf 65}, 21 (1990) [Erratum ibid. {\bf 65}, 2920 (1990)].

\bibitem{gunion}
J. F. Gunion and D. Wyler, Phys. Lett. B {\bf 248}, 170 (1990).

\bibitem{gunion2}
J. F. Gunion and R. Vega, Phys. Lett. B {\bf 251}, 157 (1990).

\bibitem{leigh}
R. G. Leigh, S. Paban, and R.-M. Xu, Nucl. Phys. B {\bf 352}, 45 (1991).

\bibitem{completebarr-zee}
D. Chang, W.-Y. Keung, and T. C. Yuan, Phys. Rev. D {\bf 43}, R14 (1991).

\bibitem{muonbarr-zee}
V. Barger, A. Das, and C. Kao, Phys. Rev. D {\bf 55}, 7099 (1997).

\bibitem{buras2hdmedm}
A. J. Buras, G. Isidori, and P. Paradisi, Phys. Lett. B {\bf 694}, 402 (2011).

\bibitem{brod}
J. Brod, U. Haisch, and J. Zupan, J. High Energy Phys. {\bf 1311} (2013) 180.

\bibitem{tabe}
T. Abe, J. Hisano, T. Kitahara, and K. Tobioka, J. High Energy Phys. {\bf 1401} (2014) 106.

\bibitem{inoue2hdmedm}
S. Inoue, M. J. Ramsey-Musolf, and Y. Zhang, Phys. Rev. D {\bf 89}, 115023 (2014). 

\bibitem{higgcisionedm}
K. Cheung, J. S. Lee, E. Senaha, and P.-Y. Tseng, J. High Energy Phys. {\bf 1406} (2014) 149. 

\bibitem{dekens}
W. Dekens, J. de Vries, J. Bsaisou, W. Bernreuther, C. Hanhart, U.-G. Mei{\ss}ner, A. Nogga, and A. Wirzba, J. High Energy Phys. {\bf 1407} (2014) 069. 

\bibitem{shuminzhao}
S.-M. Zhao, T.-F. Feng, X.-J. Zhan, H.-B. Zhang, and B. Yan, J. High Energy Phys. {\bf 1507} (2015) 124.

\bibitem{Kobakhidze}
A. Kobakhidze, N. Liu, L. Wu, and J. Yue, Phys. Rev. D {\bf 95}, 015016 (2017).

\bibitem{balazs}
C. Balazs, G. White, and J. Yue, J. High Energy Phys. {\bf 1703} (2017) 030.

\bibitem{nakai}
Y. Nakai and M. Reece, J. High Energy Phys. {\bf 1708} (2017) 031. 

\bibitem{haolinli}
C.-Y. Chen, H.-L. Li, and M. Ramsey-Musolf, Phys. Rev. D {\bf 97}, 015020 (2018).

\bibitem{Panico}
G. Panico, M. Riembau, and T. Vantalon, J. High Energy Phys. {\bf 1806} (2018) 056.

\bibitem{Bruggisser1}
S. Bruggisser, B. von Harling, O. Matsedonskyi, and G. Servant, Phys. Rev. Lett. {\bf 121}, 131801 (2018).

\bibitem{Huang}
F. P. Huang, Z. Qian, and M. Zhang, Phys. Rev. D {\bf 98}, 015014 (2018).

\bibitem{Bruggisser2}
S. Bruggisser, B. von Harling, O. Matsedonskyi, and G. Servant, J. High Energy Phys. {\bf 1812}, 099 (2018).

\bibitem{mssm1}
H. E. Haber and G. L. Kane, Phys. Rep. {\bf 117}, 75 (1985).

\bibitem{mssm2}
J. F. Gunion and H. E. Haber, Nucl. Phys. B {\bf 272}, 1 (1986).

\bibitem{mssm3}
S. P. Martin, in {\it Perspectives on Supersymmetry II}, edited by G. L. Kane (World Scientific, Singapore, 2010), p. 1, Adv. Ser. Direct. High Energy Phys. {\bf 21}, 1 (2010); {\bf 18}, 1 (1998).

\bibitem{bingli}
B. Li and C. E. M. Wagner, Phys. Rev. D {\bf 91}, 095019 (2015).

\bibitem{carena}
M. Carena, J. Ellis, J. S. Lee, A. Pilaftsis, and C. E. M. Wagner, J. High Energy Phys. {\bf 1602} (2016) 123.

\bibitem{Heinemeyer1}
S. Heinemeyer and C. Schappacher, Eur. Phys. J. C {\bf 76}, 535 (2016).

\bibitem{passehr}
S. Pa{\ss}ehr and G. Weiglein, Eur. Phys. J. C {\bf 78}, 222 (2018).

\bibitem{borowka}
S. Borowka, S. Pa{\ss}ehr, and G. Weiglein, Eur. Phys. J. C {\bf 78}, 576 (2018).

\bibitem{Heinemeyer2}
S. Heinemeyer and C. Schappacher, Eur. Phys. J. C {\bf 78}, 536 (2018).

\bibitem{Chang:1998uc} 
D.~Chang, W.~Y.~Keung, and A.~Pilaftsis, Phys.\ Rev.\ Lett.\  {\bf 82}, 900 (1999) Erratum: [Phys.\ Rev.\ Lett.\  {\bf 83}, 3972 (1999)].

\bibitem{Pilaftsis:1999td} 
A.~Pilaftsis, Phys.\ Lett.\ B {\bf 471}, 174 (1999).

\bibitem{Chang:1999zw} 
D.~Chang, W.~F.~Chang, and W.~Y.~Keung, Phys.\ Lett.\ B {\bf 478}, 239 (2000).

\bibitem{Pilaftsis:2002fe} 
A.~Pilaftsis, Nucl.\ Phys.\ B {\bf 644}, 263 (2002).

\bibitem{Yamanaka:2012ia} 
N.~Yamanaka, Phys.\ Rev.\ D {\bf 87}, 011701 (2013).

\bibitem{Sirunyan:2017leh} 
A.~M.~Sirunyan {\it et al.} (CMS Collaboration), Phys.\ Rev.\ D {\bf 97}, 032009 (2018).

\bibitem{Aaboud:2018kya} 
M.~Aaboud {\it et al.} (ATLAS Collaboration), Phys.\ Rev.\ D {\bf 98}, 032008 (2018).

\bibitem{Aaboud:2018mna} 
M.~Aaboud {\it et al.} (ATLAS Collaboration), Phys. Rev. D {\bf 99}, 012009 (2019).

\bibitem{cancellationhiggs1}
L. Bian, T. Liu, and J. Shu, Phys. Rev. Lett. {\bf 115}, 021801 (2015).

\bibitem{cancellationhiggs2}
L. Bian and N. Chen, Phys. Rev. D {\bf 95}, 115029 (2017).

\bibitem{rpv1}
G. Bhattacharyya, arXiv:hep-ph/9709395.

\bibitem{rpv2}
H. K. Dreiner, in {\it Perspectives on Supersymmetry II}, edited by G. L. Kane (World Scientific, Singapore, 1997), p. 565, Adv. Ser. Direct. High Energy Phys. {\bf 21}, 565 (2010).

\bibitem{rpv3}
M. Chemtob, Prog. Part. Nucl. Phys. {\bf 54}, 71 (2005).

\bibitem{rpv4}
R. Barbier {\it et al.}, Phys. Rep. {\bf 420}, 1 (2005).

\bibitem{Aaboud:2018zeb} 
M.~Aaboud {\it et al.} (ATLAS Collaboration), Phys.\ Rev.\ D {\bf 98}, 032009 (2018).

\bibitem{rpvbarr-zee1}
N. Yamanaka, T. Sato, and T. Kubota, Phys. Rev. D {\bf 85}, 117701 (2012).

\bibitem{rpvbarr-zee2}
N. Yamanaka, T. Sato, and T. Kubota, Phys. Rev. D {\bf 87}, 115011 (2013).

\bibitem{leptoquark1}
S. Davidson, D. Bailey, and B. A. Campbell, Z. Phys. C {\bf 61}, 613 (1994).

\bibitem{leptoquark2}
J. L. Hewett and T. G. Rizzo, Phys. Rev. D {\bf 56}, 5709 (1997).
  
\bibitem{leptoquark3}
P. Nath and P. Fileviez Perez, Phys. Rep.{\bf 441}, 191 (2007).

\bibitem{leptoquark4}
I. Dor\u{s}ner, S. Fajfer, A. Greljo, J. F. Kamenik, and N. Ko\u{s}nik,  Phys. Rep. {\bf 641}, 1 (2016).

\bibitem{Aaboud:2016qeg} 
M.~Aaboud {\it et al.} (ATLAS Collaboration), New J.\ Phys.\  {\bf 18}, 093016 (2016).

\bibitem{Sirunyan:2018ryt} 
A.~M.~Sirunyan {\it et al.} (CMS Collaboration), Phys. Rev. D {\bf 99}, 032014 (2019).

\bibitem{Sirunyan:2018ruf} 
A.~M.~Sirunyan {\it et al.} (CMS Collaboration), Phys. Rev. Lett. {\bf 121}, 241802 (2018).

\bibitem{Sirunyan:2018btu} 
A.~M.~Sirunyan {\it et al.} (CMS Collaboration), Phys. Rev. D {\bf 99}, 052002 (2019).

\bibitem{Barr:1987sp} 
S.~M.~Barr and A.~Masiero, Phys. Rev. Lett.  {\bf 58}, 187 (1987).

\bibitem{Geng:1990gr} 
C.~Q.~Geng, Z.\ Phys.\ C {\bf 48}, 279 (1990).

\bibitem{pospelovsmatomicedm}
M. Pospelov and A. Ritz, Phys. Rev. D {\bf 89}, 056006 (2014).

\bibitem{yamanakasmedm}
N. Yamanaka and E. Hiyama, J. High Energy Phys. {\bf 1602} (2016) 067.

\bibitem{yamanakasmedm2}
N. Yamanaka, Nucl. Phys. A {\bf 963}, 33 (2017).

\end{thebibliography}
\end{document}